**Thermodynamic evidence for interaction-driven first-order topological quantum phase transitions**

Surajit Dutta[1†], Nadav Auerbach[1†], Chiho Yoon[2†], Tonghang Han[3†], Matan Uzan[1], Zhengguang Lu[4], Niladri-Sekhar Kander[1], Yaozhang Zhou[1], Yuri Myasoedov[1], Martin E. Huber[5], Kenji Watanabe[6], Takashi Taniguchi[7], Long Ju[3], Fan Zhang[2], and Eli Zeldov[1]*

Topological quantum phase transitions in non-interacting systems occur through continuous gap closing and reopening. In strongly interacting systems, however, competing ordered states have long been predicted to drive first-order transitions, although this possibility has remained experimentally unresolved. Recent transport studies of correlated phases in charge-neutral rhombohedral graphene were interpreted as evidence for continuous topological transitions. Here, using nanoSQUID-on-tip magnetometry, we directly image the local orbital magnetization of a spin–orbit-proximitized rhombohedral graphene quantum anomalous Hall (QAH) state. We provide the first real-space visualization of a QAH phase with a record Chern number, reconstruct its local thermodynamic gap, and track the evolution of its magnetization across competing correlated states. Combined with self-consistent Hartree–Fock calculations, these measurements show that the sequential transitions between the layer-antiferromagnetic, QAH, and layer-polarized insulating states are first order, accompanied by discontinuous changes in orbital magnetization. Near the phase boundaries, we observe fluctuating magnetic domains, providing direct microscopic evidence of phase coexistence between nearly degenerate competing ordered states. Together, these observations provide the first direct thermodynamic evidence for first-order topological quantum phase transitions and establish a microscopic framework for understanding interaction-driven topological quantum phase transitions through phase competition and coexistence.

[1]Department of Condensed Matter Physics, Weizmann Institute of Science, Rehovot 7610001, Israel
[2]Department of Physics, University of Texas at Dallas, Richardson, TX, USA
[3]Department of Physics, Massachusetts Institute of Technology, Cambridge, MA, USA
[4]Department of Physics, Florida State University, Tallahassee, FL, USA
[5]Departments of Physics and Electrical Engineering, University of Colorado Denver; Denver, Colorado 80217, USA
[6]Research Center for Electronic and Optical Materials, National Institute for Materials Science; 1-1 Namiki, Tsukuba 305-0044, Japan
[7]Research Center for Materials Nanoarchitectonics, National Institute for Materials Science; 1-1 Namiki, Tsukuba 305-0044, Japan
[†]These authors contributed equally to this work
*eli.zeldov@weizmann.ac.il

Electronic topological phases are distinguished by quantized topological invariants that remain unchanged under continuous deformation of the electronic structure. Consequently, in non-interacting systems, a change in the topological invariant requires the continuous closing and reopening of a bulk energy gap [1–3]. In strongly interacting systems, however, competition between nearly degenerate ordered states has long been predicted to fundamentally alter this paradigm by driving discontinuous first-order topological phase transitions [4]. Rhombohedral graphene (RG), which hosts both topologically trivial and nontrivial correlated insulating states at charge neutrality [5], provides an exceptional platform testing this prediction. Early theoretical works proposed that the electron-electron interactions and displacement field induce first-order topological transitions [6–8] accompanied by phase coexistence, domain formation, conducting domain walls, and fluctuations [9,10]. Yet direct experimental evidence capable of distinguishing between these two fundamentally different scenarios has remained elusive. Instead, recent transport studies of tetralayer and pentalayer RG interpreted the interaction- and field-driven transitions as continuous topological phase transitions [11,12].

RG intrinsically hosts two low-energy flat bands localized on its outermost layers and near the $K$ and $K'$ corners of the Brillouin zone, where the Berry curvature is sharply peaked. Consequently, electronic interactions stabilize a rich landscape of correlated phases [5,13–16]. These include isospin magnets [17–21], correlated insulators [19–25], integer and fractional quantum anomalous Hall (QAH) states [11,12,25–29], multiferroics [21,25,30–33], superconductors [21,25,34–38], and Wigner crystals [39]. At charge neutrality, the dominant competing phases are the layer antiferromagnetic (LAF) and layer-polarized insulating (LPI) states [19–21]. When sufficiently strong proximity-induced spin–orbit coupling is introduced, a QAH state emerges between them [11,12]. A defining feature of this family of broken-symmetry phases is that their Chern number, orbital magnetization, and electric polarization are all governed by the degree of layer polarization [5]. Consequently, an applied displacement field simultaneously tunes the topological character, orbital magnetization, and energetic competition between the three phases, making RG an ideal platform for investigating interaction-driven topological quantum phase transitions.

Determining the order of a topological phase transition requires measuring a thermodynamic quantity. While transport measurements determine the Chern number of a gapped state through quantized Hall conductance, they cannot determine the order of the transition itself. Magnetization, by contrast, is a derivative of free energy with respect to magnetic field and therefore provides a direct thermodynamic probe. For the RG QAH state, orbital magnetization also serves as its natural order parameter. Moreover, orbital magnetization is sensitive to chemical potential [40–42] and thus evolves in both Chern insulating and metallic regimes. Together, these properties render orbital magnetization a thermodynamic lens into the microscopic evolution of interaction-driven competing phases across topological transitions.

Here, using ultrasensitive SQUID-on-tip magnetometry [18,43–45], we directly measure the local orbital magnetization of spin–orbit-proximitized RG pentalayer. We provide the first real-space visualization of a $C = \pm 5$ QAH state, revealing its magnetic texture and reconstructing its local thermodynamic gap. By combining local magnetic imaging with self-consistent Hartree–Fock calculations, we determine the thermodynamic nature of the quantum phase transitions between the competing LAF, QAH, and LPI states. These measurements provide the first direct thermodynamic evidence for first-order topological quantum phase transitions in strongly correlated electron systems, revealing a microscopic picture for interaction-driven topological transitions undergoing phase competition and coexistence between nearly degenerate ordered states, rather than the conventional paradigm of continuous evolution from one phase to another.

**Transport measurements**

The $WS_2$-proximitized rhombohedral pentalayer graphene (R5G) device was fabricated in a Hall bar geometry with dual graphite gates, as illustrated in Fig. 1d. To avoid moiré superlattice effects, the R5G was intentionally misaligned with both the bottom hBN and the top $WS_2$ layers. The *dc* voltages applied to the top and bottom gates, $V_{tg}^{dc}$ and $V_{bg}^{dc}$, allow for independent control of the carrier density, $n = (C_{tg}V_{tg}^{dc} + C_{bg}V_{bg}^{dc})/e$, and the transverse displacement field, $D = \frac{1}{2\varepsilon_0}(C_{tg}V_{tg}^{dc} - C_{bg}V_{bg}^{dc})$, where $\varepsilon_0$ is the vacuum permittivity and $C_{tg}$ and $C_{bg}$ are the top and bottom capacitances per unit area. Figure 1a shows the longitudinal resistance $R_{xx}(n, D)$ over a wide range of $n$ and $D$ at zero applied magnetic field, $B_a = 0$ T and base temperature $T = 20$ mK. High resolution maps of $R_{xx}(n, D)$ and $R_{yx}(n, D)$ near charge neutrality, corresponding to the dashed rectangle in Fig. 1a, are shown in Figs. 1b,c at $B_a = 12$ mT (see Extended Data Figs. 1a-c for additional fields).

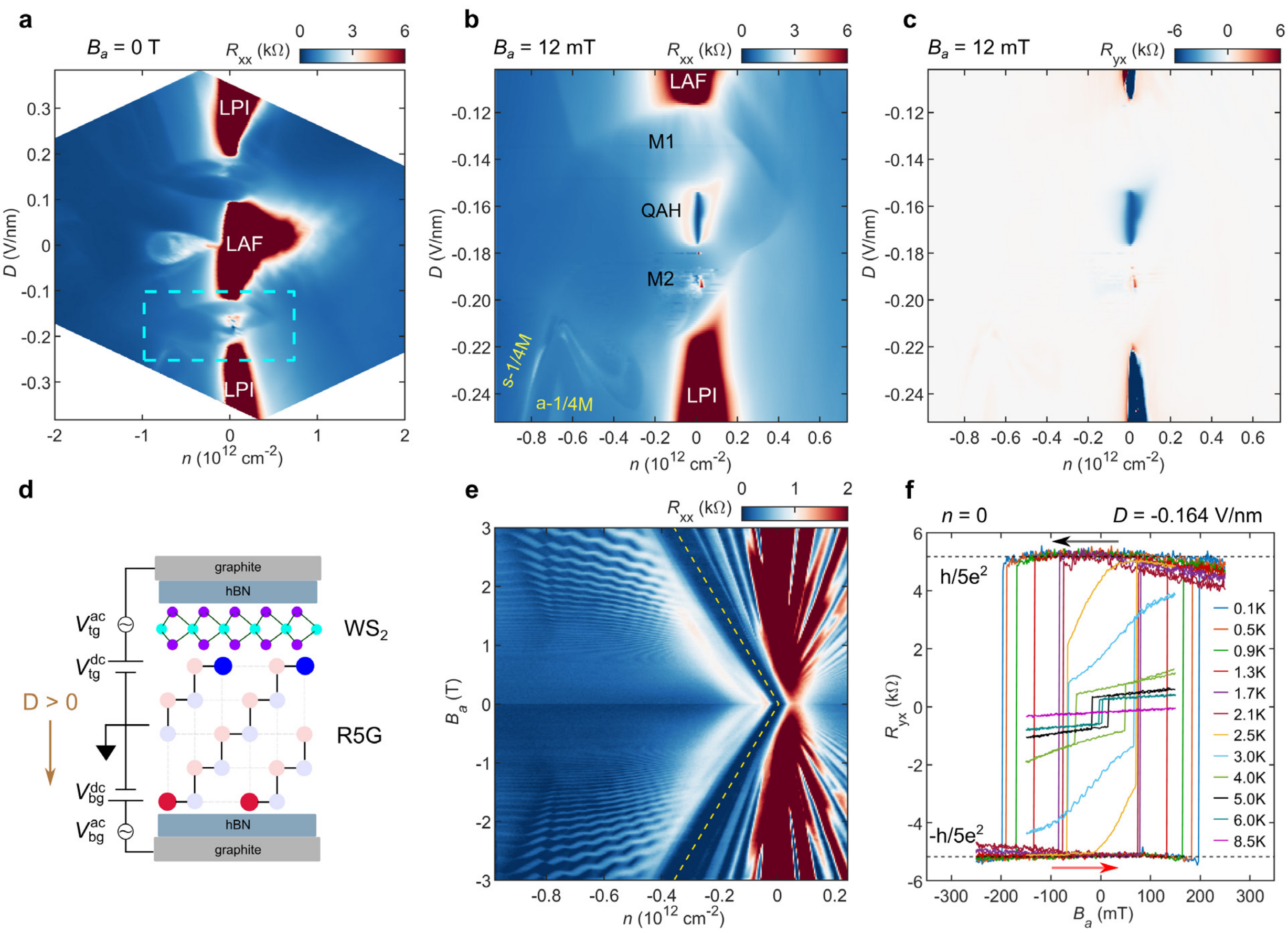

**Fig. 1. Transport measurements of $WS_2$-proximitized rhombohedral pentalayer graphene. a,** Longitudinal resistance $R_{xx}(n, D)$ measured in zero magnetic field $B_a$ at $T = 20$ mK. **b**,**c**, High-resolution maps of $R_{xx}$ (**b**) and Hall resistance $R_{yx}$ (**c**) in the dashed cyan box in (a). **d,** Schematics of the R5G proximitized to $WS_2$ device and of the electrical circuit. **e,** $R_{xx}(n, B_a)$ at $D = -0.164$ V/nm at $T = 100$ mK. The yellow dotted lines represent the Streda formula fits with $C = \pm 5$. **f,** $R_{yx}$ as a function of sweeping $B_a$ up (red arrow) and down (black arrow) in the QAH state at $D = -0.164$ V/nm at different indicated temperatures.

At charge neutrality ($n = 0$), the R5G device exhibits a sequence of distinct phases as a function of the displacement field. For $|D| \lesssim 0.12$ V/nm, the system displays an insulating state with $R_{xx} \sim 1$ MΩ, attributed to a LAF phase [5,19–24,26,27]. This correlated insulator arises from strong Coulomb interactions in the nearly flat touching bands and is characterized by opposite top-or-bottom layer polarizations of the opposite spins. At large displacement fields, $|D| \gtrsim 0.22$ V/nm, the system transitions into a LPI phase, where $R_{xx}$ increases monotonically with $|D|$, consistent with a single-particle gap that grows with the displacement potential.

Between the LAF and LPI phases, distinct metallic-like regions emerge at intermediate $|D|$. The metallic-like region at $D > 0$ shows a rather uniform behavior at $B_a = 0$ T (Fig. 1a). In contrast, at negative $D$ a more complex structure featuring three regions occurs: two metallic-like states, labeled M1 and M2, which flank a central QAH state (Figs. 1b,c). This QAH state, observed over the range $-0.177 < D < -0.154$ V/nm, has a quantized $R_{yx} = \pm\frac{h}{5e^2}$ (Fig. 1f) with vanishing $R_{xx}$ at charge neutrality at zero magnetic field (Fig. 2f). Moreover, the M2 region shows pronounced fluctuations in $R_{xx}$ and $R_{yx}$ (Figs. 1b,c).

The Landau fan diagram measured at $D = -0.164$ V/nm (Fig. 1e) reveals a dominant Landau level (LL) emanating from $n = 0$ at $B_a = 0$ T, with a slope corresponding to a Chern number $C = -5$ for $B_a > 0$. Additional electron- or hole-like LLs appear only at higher fields, above 1 T. Figure 1f shows the anomalous $R_{yx}$ measured at different temperatures. At the lowest $T$, $R_{yx}$ exhibits magnetic hysteresis, switching between the two quantized values $\pm\frac{h}{5e^2}$, providing compelling evidence that the system hosts a QAH state with $|C| = 5$. The anomalous Hall effect persists up to 8.5 K, defining the Curie temperature of the ferromagnetic order.

**Imaging QAH magnetism**

To visualize the magnetic texture and evolution of the QAH state, we performed nanoscale magnetic imaging using a scanning superconducting quantum interference device (SQUID-on-tip, SOT), fabricated at the apex of a sharp quartz pipette [43–45]. In addition to the *dc* gate voltages ($V_{tg}^{dc}$, $V_{bg}^{dc}$), small *ac* voltages ($V_{tg}^{ac}$, $V_{bg}^{ac}$) were superimposed to modulate $n$ and $D$ by $n^{ac}$ and $D^{ac}$. These modulations induce an *ac* change in the local magnetization, which in turn produces an *ac* stray magnetic field $B_z^{ac}(x, y)$. Measured by the SOT, $B_z^{ac}$ provides a direct probe of the local differential magnetization [18,45–47]. Figure 2b shows the $n$-$D$ map of $B_z^{ac}$ measured at a single spatial point in the bulk of the sample. At finite hole densities ($n = -0.9$ to $-0.4\times10^{10}$ cm$^{-2}$), the transition lines associated with the simple and annular quarter metal (s-1/4M and a-1/4M) phases—observable also in transport (Fig. 1b and Extended Data Fig. 1) —are clearly resolved in the magnetic signal.

Next, we focus on the magnetic response of the QAH state. Near $n = 0$, clear magnetic features are observed within the displacement field range corresponding to the QAH phase identified in transport (Fig. 1b). To resolve these features more precisely, we measured $B_z^{ac}$ over a much narrower window in $n$ and $D$, using a small top gate modulation, $V_{tg}^{ac} = 6$ mV (rms). The resulting map (Fig. 2c) reveals a distinctive tri-striped pattern: a central negative signal near $n = 0$, flanked by positive signals on both sides. To understand the origin of this unusual pattern, we calculate the orbital magnetization $\mathcal{M}$ as a function of the chemical potential $\mu$ using a self-consistent Hartree-Fock (scHF) theory (Fig. 2d and Methods). The orbital magnetization arising from Berry curvature has two contributions [5,6,40–42,45] (Methods): a self-rotation magnetization term $\mathcal{M}_{SR}$ and the Chern magnetization term $\mathcal{M}_C$,. Within the Chern gap, $\mathcal{M}_{SR}$ remains constant, while $\mathcal{M}_C$ varies linearly with $\mu$, with a universal slope $\frac{\partial \mathcal{M}_C}{\partial \mu} = C\frac{e}{h}$ [41,45]. For the $C = -5$ QAH state, this yields a negative $\frac{\partial \mathcal{M}}{\partial \mu}$ near $\mu = 0$ (Fig. 2d). The central negative $B_z^{ac}$ signal in Fig. 2c directly reflects this negative differential magnetization in the gapped QAH state. Outside the gap, the magnetization behavior is dominated by $\mathcal{M}_{SR}$, resulting in positive $\frac{\partial \mathcal{M}}{\partial \mu}$ and correspondingly positive $B_z^{ac}$ signal—completing the tri-striped pattern. While the scHF theory successfully reproduces the overall tri-striped pattern, it does not capture the asymmetry, with a weaker $B_z^{ac}$ response on the electron-doped side. The origin of this asymmetry likely reflects more detailed band-edge renormalization beyond the present scHF theory, which primarily addresses spontaneous gap opening.

Similar $B_z^{ac}$ maps were acquired at various $B_a$ and compiled onto the $n$-$B_a$ plane as shown in Fig. 2e. The centers of the maps closely follow the Streda formula lines corresponding to $C = \pm5$, as determined from transport (dotted lines in Fig. 1e). Notably, for negative $B_a$, the system favors a different QAH state with the opposite $C$, resulting in reversed differential magnetization and colors in the tri-striped patterns.

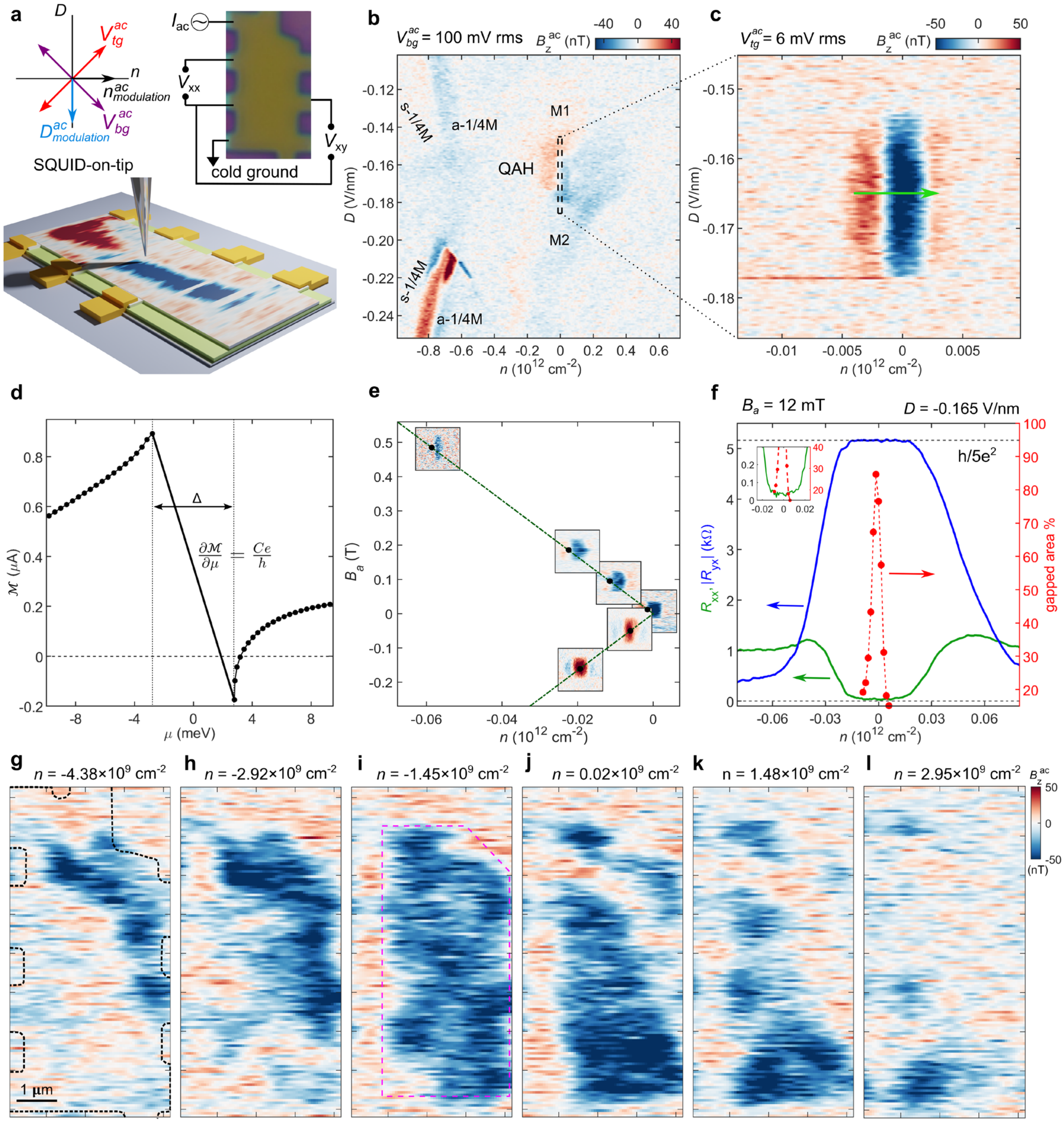

**Fig. 2. Orbital magnetism in the QAH state. a,** Schematics of the scanning SOT imaging setup (bottom), optical image of the device (top-right), and schematics of the different modulation schemes $V_{tg}^{ac}$, $V_{bg}^{ac}$, $n^{ac}$, and $D^{ac}$ used for differential magnetic imaging (top-left). **b,** $B_z^{ac}(n, D)$ measured by the SOT at a single location in the sample at $B_a$ = 12 mT and $T$ = 20 mK using $V_{bg}^{ac}$ = 100 mV rms. **c,** Zoomed-in map of $B_z^{ac}(n, D)$ in the vicinity of QAH state within the dashed box in (b) using $V_{tg}^{ac}$ = 6 mV rms. **d,** Calculated total magnetization $\mathcal{M}(\mu)$ within the scHF theory as a function of chemical potential $\mu$ in the QAH state. Within the topological gap Δ, the orbital magnetization has a universal slope $\frac{\partial \mathcal{M}}{\partial \mu} = C\frac{e}{h}$. **e,** Maps of $B_z^{ac}(n, D)$ at several $B_a$ values. For $B_a > 0$ ($B_a < 0$) the centers of the maps follow $C = -5$ ($C = 5$) Streda lines (dotted) with correspondingly negative (positive) $B_z^{ac}$ in the gapped QAH regions. **f,** Left axis: linecut of $R_{xx}$ (green) and $R_{yx}$ (blue) vs. $n$ measured at $D = -0.165$ V/nm and $B_a$ =12 mT. Right axis: percentage of sample area that is in the gapped QAH state within the magenta dashed box in (i), calculated from (g)-(l). The red dashed line is a guide to the eye. Inset: zoomed-in view of $R_{xx}$ and of the gapped-area percentage near charge neutrality. **g-l,** $B_z^{ac}(x, y)$ maps acquired at a constant $D = -0.165$ V/nm and variable $n$ along the green arrow in (c) at $B_a = 12$ mT and $V_{bg}^{ac} = 6$ mV rms. See Extended Data Fig. 2 for similar $B_z^{ac}(x,y)$ maps at fixed $n$ and varying $D$. The black dashed frame in (g) outlines the sample boundaries.

To directly visualize the spatial distribution of orbital magnetization in the QAH state, Figs. 2g-l show two-dimensional $B_z^{ac}(x,y)$ images acquired at fixed $D = -0.165$ V/nm and varying carrier density along the green arrow in Fig. 2c. Upon approaching charge neutrality from the hole-doped side, domains with negative differential magnetization first appear near the top-right edge of the sample (Fig. 2g), indicating regions that have entered the gapped QAH state. As the hole doping is further reduced, the gapped regions expand toward the center of the sample (Fig. 2h). Near charge neutrality (Figs. 2i,j), most of the sample area enters the gapped QAH state, before shrinking again upon electron doping. This spatial evolution highlights the distinction between local magnetization and global transport. Although the QAH state is fully gapped wherever it forms, different regions enter the gap at different nominal carrier densities due to charge disorder. Thus, a locally abrupt onset of the QAH gap can appear as a continuous evolution of the transport activation gap in spatially averaged Arrhenius analyses [11,12].

It is important to note that the nominal carrier density, calculated from $n = (C_{tg}V_{tg}^{dc} + C_{bg}V_{bg}^{dc})/e$, represents a spatial average. Locally, however, the carrier density has substantial spatial variations arising from charge disorder in the hBN and graphite gates. Hence at $n \cong 0$, some regions, like the bottom-left corner in Fig. 2i, have not yet reached the gap and are hole doped, while other regions, like the tilted red strip in the upper part of the sample in Fig. 2j, have already passed the gap and are electron doped. As the electron doping is further increased (Figs. 2k), the gapped areas shrink, until most of the sample becomes metallic again (Fig. 2l). Note that the 2D images were acquired at positive field $B_a = 12$ mT at which the $C = -5$ QAH state is stabilized. As a result, only domains with negative differential magnetization are observed; domains with positive differential magnetization corresponding to the $C = 5$ QAH state—stabilized under negative $B_a$—are absent.

To quantify the evolution of the gapped QAH domains with doping, we determine the fraction of the sample area in the gapped QAH state (blue areas in Figs. 2g-l with $B_z^{ac} < 0$) within the magenta dashed box in Fig. 2i. As shown by the red symbols in Fig. 2f, the QAH area fraction peaks sharply at $n = 0$, where most of the sample is gapped, and rapidly decreases with either electron or hole doping. In an ideal disorder-free system, this peak would be infinitely sharp, limited only by the carrier modulation amplitude $n^{ac} = 3.45\times10^9$ cm$^{-2}$. The observed broader peak reflects the degree of charge disorder in the device, which we evaluate from its full width at half maximum to be $\delta n \cong 5.9\times10^9$ cm$^{-2}$. This $\delta n$ is an order of magnitude lower than values reported in MATBG [43], highlighting the superior cleanliness of the crystalline rhombohedral graphene compared to moiré graphene. Notably, $R_{xx}$ increases rapidly as the gapped regions shrink and the sample transitions into the metallic state, as highlighted in the inset of Fig. 2f near charge neutrality. In contrast, $R_{yx}$ remains close to its quantized value over a broader density range because the Hall conductivity is tied to the underlying Chern topology and is therefore more robust against doping and dissipative transport.

Additional support for charge disorder as the dominant source of inhomogeneity comes from Extended Data Fig. 2, which shows the $B_z^{ac}(x,y)$ maps acquired while sweeping $D$ at fixed $n \cong -0.96\times10^9$ cm$^{-2}$. In contrast to the evolving patterns observed by varying $n$ (Figs. 2g-l), the spatial pattern remains essentially static as $D$ is tuned within the QAH regime. This insensitivity to $D$ further confirms that local doping inhomogeneities—not displacement field variations [48]—dominate the disorder landscape in the device.

The differential magnetization enables direct determination of the thermodynamic gap of the QAH state. As illustrated by the calculated orbital magnetization $\mathcal{M}(n)$ in Fig. 3a, there emerges a sharp step $\delta\mathcal{M}$ across the QAH gap, followed by a much slower evolution of $\mathcal{M}$ with carrier density outside the gap. Consequently, a small carrier-density modulation primarily measures the step $\delta\mathcal{M}$, with only weak dependence on the modulation amplitude. We reconstruct the local magnetization jump, $\delta\mathcal{M}(x,y)$, from the measured $B_z^{ac}$ map in Fig. 2i by numerically inverting the stray magnetic field (Methods). The resulting map (Fig. 3b) directly provides the thermodynamic gap map through the universal relation $\delta\mathcal{M} = C(e/h)\Delta$ (Fig. 3c). To validate the reconstruction, we calculate the stray field generated by the reconstructed magnetization using the Biot–Savart law. The resulting $B_z^{ac}$ map (Fig. 3d) quantitatively reproduces the measured signal in Fig. 2i, demonstrating the accuracy of the inversion procedure.

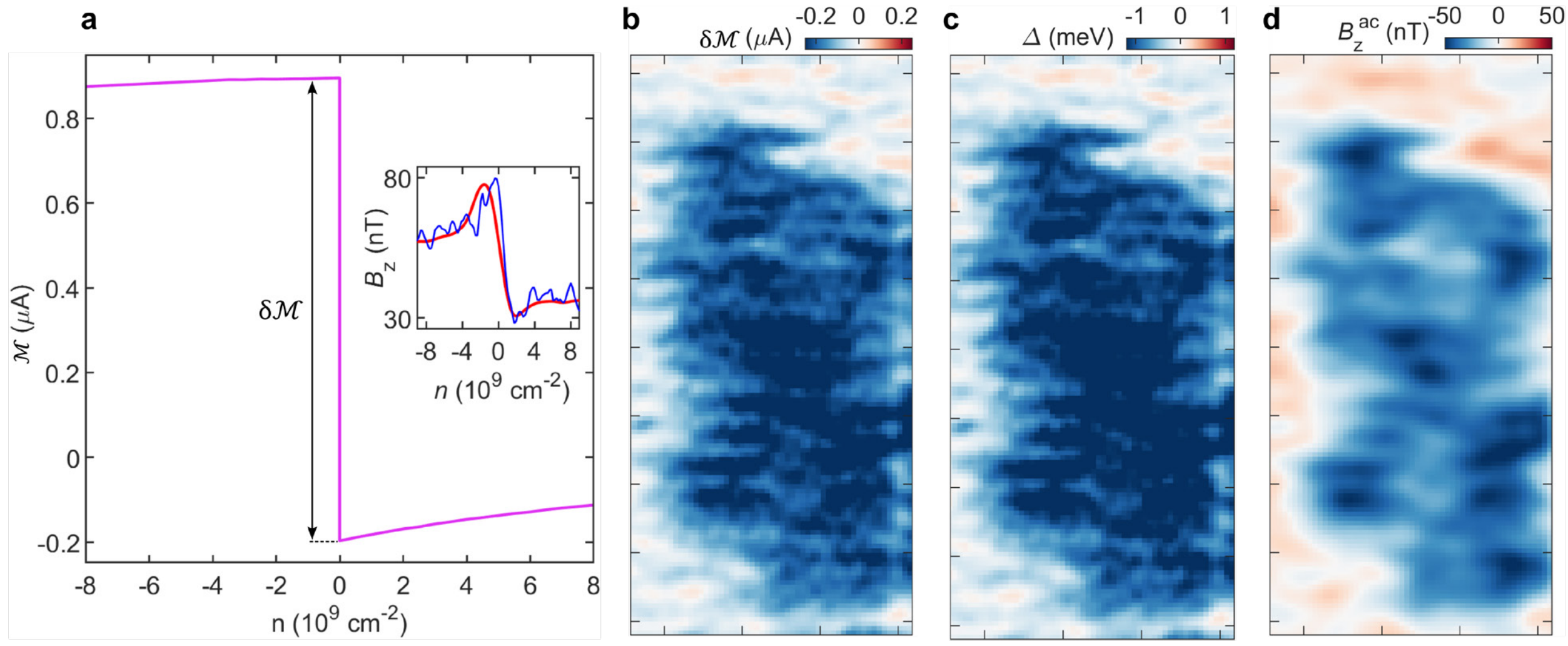

**Fig. 3. Local magnetization reconstruction and thermodynamic gap of QAH state. a,** Calculated orbital magnetization $\mathcal{M}(n)$ as a function of $n$ in the QAH state within the scHF theory (equivalent to $\mathcal{M}(\mu)$ in Fig. 2d), showing a sharp jump $\delta\mathcal{M} = C\frac{e}{h}\Delta$ across the topological gap. Inset: Red curve, integrated magnetic signal $B_z(n) \propto \mathcal{M}(n)$, obtained by integrating the $B_z^{ac}(n) \propto \partial\mathcal{M}(n)/\partial n$ data in Fig. 4a along the $n$ direction and averaging over the $D$ range of the QAH state. Blue curve, amplitude of the $D^{ac}$-modulated magnetic response, $|B_z^{ac}|$, extracted along the $D_{c2}$ transition line vs. $n$ in Fig. 4b and rescaled to the red curve. **b,** Reconstructed magnetization jump, $\delta\mathcal{M}(x,y)$, across the QAH gap, obtained by numerically inverting the measured $B_z^{ac}(x,y)$ map in Fig. 2i. **c,** Spatial map of the topological gap $\Delta(x,y)$ calculated from (b) using the universal relation $\delta\mathcal{M} = C\frac{e}{h}\Delta$. **d,** $B_z^{ac}(x,y)$ calculated from the reconstructed $\delta\mathcal{M}(x,y)$ in (b) using the Biot–Savart law, quantitatively reproducing the measured $B_z^{ac}(x,y)$ map in Fig. 2i, validating the inversion procedure.

The $\Delta(x,y)$ map in Fig. 3c reveals that the QAH gap is rather uniform through the bulk of the sample. This indicates that the finite width of the global QAH transition is governed primarily by charge disorder, which shifts the local carrier density at which different regions enter the QAH state, rather than by significant variations in the gap magnitude itself. Averaging over the reconstructed map yields a QAH gap of $\Delta \approx 1.2$ meV ($\approx 14$ K), which compares well with the Curie temperature of 8.5 K extracted independently from the magnetic hysteresis measurements (Fig. 1f). The experimentally determined gap is, however, smaller than the $\Delta \approx 5.6$ meV predicted by the scHF calculations (Fig. 2d), consistent with the well-known tendency of mean-field theories to overestimate interaction-induced gaps owing to the neglect of quantum fluctuations and correlation effects.

### First-order topological transitions

Next, we turn to the emergence of the QAH state and the thermodynamic nature of its transitions into the neighboring correlated phases. To distinguish between a first-order (FO) topological transition (Fig. 4h) and a continuous topological transition (Fig. 4i), we directly measure $B_z^{ac}$ using either pure carrier-density modulation $n^{ac}$ or pure displacement-field modulation $D^{ac}$ (Fig. 2a schematic). Under $n^{ac}$ modulation (Fig. 4a), the tri-striped magnetic pattern observed previously in Fig. 2c persists. Strikingly, under $D^{ac}$ modulation, the tri-striped pattern vanishes and is replaced by two sharp peaks of opposite polarity at the critical displacement fields $D_{c1} = -0.154$ V/nm and $D_{c2} = -0.178$ V/nm (Fig. 4b and Extended Data Fig. 4). For this modulation scheme, $B_z^{ac}$ is proportional to the differential orbital magnetization $\partial\mathcal{M}/\partial D$; thus, the two sharp peaks correspond to discontinuous steps in $\mathcal{M}(D)$. Since the orbital magnetization is a thermodynamic quantity $\mathcal{M} = -\partial F/\partial B$, these discontinuities in a first derivative of the free energy $F$ provide direct thermodynamic evidence for FO transitions between the QAH state and the competing correlated states and are inconsistent with a continuous transition, which would instead produce a smooth evolution of the $\mathcal{M}(D)$ (green curve in Fig. 4g, Extended Data Fig. 5f, and Supplementary Video 2).

To understand these features, we performed scHF calculations (Methods). Figure 4d shows the total energies $E_{\text{tot}}$ of the competing LAF, QAH, and LPI states at charge neutrality. At small negative displacement potential $U$, the LAF state (blue line) exhibits the lowest energy with up-spin conduction bands ($K\uparrow$ and $K'\uparrow$) polarized to top layer with positive gaps, while down-spin bands ($K\downarrow$ and $K'\downarrow$) are polarized to bottom layer carrying negative gaps (Fig. 4h). This minimizes the Hartree energy by dictating no overall layer polarization while gaining both intra-valley and valley-interchange exchange energies [6]. Absent competition, the LAF state would persist up to $U \approx -11$ meV, where the $K\downarrow$ and $K'\downarrow$ gaps close (Extended Data Figs. 5d,e). However, the QAH state (red line), although higher in energy at low $|U|$, becomes progressively more favored with increasing $|U|$ due to its net layer polarization. At $U_{c1} = -6.81$ meV, its $E_{tot}$ crosses below that of the LAF state (Fig. 4d), producing a FO transition in which the $K\uparrow$ flavor abruptly reverses its layer polarization and hence its gap sign (Fig. 4h and Supplementary Video 1). At larger $|U|$, a second FO transition occurs at $U_{c2} = -7.98$ meV, where the QAH state transitions into the LPI state through a discontinuous reversal of the $K'\uparrow$ flavor layer polarization and gap sign (Fig. 4h).

The abrupt change in layer polarization produces a discontinuous jump in $\mathcal{M}$ (Fig. 4g, magenta curve). This behavior is fundamentally different from a continuous transition, where $\mathcal{M}$ evolves smoothly across the gap closing and reopening (Fig. 4g green curve and Extended Data Fig. 5f). Consequently, although both scenarios exhibit discrete changes in the total $C$, only the FO transition generates a discontinuity in $\mathcal{M}$, consistent with our observations. At the LAF–to–QAH transition, $\mathcal{M}$ jumps from nearly zero to a finite positive value, producing a sharp positive peak in differential $\mathcal{M}^{ac} \cong d\mathcal{M}/dU$ (Methods) and corresponding $B_z^{ac}$ (Fig. 4g blue curve). At the QAH–to–LPI transition, $\mathcal{M}$ drops abruptly to zero, resulting in a sharp negative peak. Notably, the experimentally observed asymmetry between the two peaks in Fig. 4b, with a substantially weaker response at $D_{c1}$, is qualitatively reproduced by the calculated $\mathcal{M}(U)$, which exhibits a smaller magnetization jump at $U_{c1}$ than at $U_{c2}$ (Fig. 4g).

Because these transitions depend almost exclusively on $D$, with only weak dependence on $n$, they are visible only under $D^{ac}$ modulation (Fig. 4b) and disappear under $n^{ac}$ modulation (Fig. 4a). With $V_{tg}^{ac}$ modulation, which simultaneously modulates $D$ and $n$, the two signatures coexist, giving rise to both the tri-striped pattern and the FO features (Fig. 2c). These two orthogonal modulation schemes provide an important consistency check of the calculated $\mathcal{M}(n)$, which exhibits a sharp step at the QAH gap followed by a gradual decay with doping (Fig. 3a). Integrating the measured $B_z^{ac} \propto \partial\mathcal{M}(n)/\partial n$ in Fig. 4a along $n$ reconstructs the functional form of $\mathcal{M}(n)$ (red curve in Fig. 3a inset), qualitatively reproducing the calculated $\mathcal{M}(n)$ broadened by the finite $n^{ac}$ modulation. Independently, the peak amplitude of the $D^{ac}$-modulated response along the $D_{c2}$ transition in Fig. 4b (blue curve), which is proportional to the magnetization step between the magnetic QAH and nearly nonmagnetic M2 states, follows the same carrier-density dependence. The close agreement between these complementary measurements further supports the FO nature of the QAH transition.

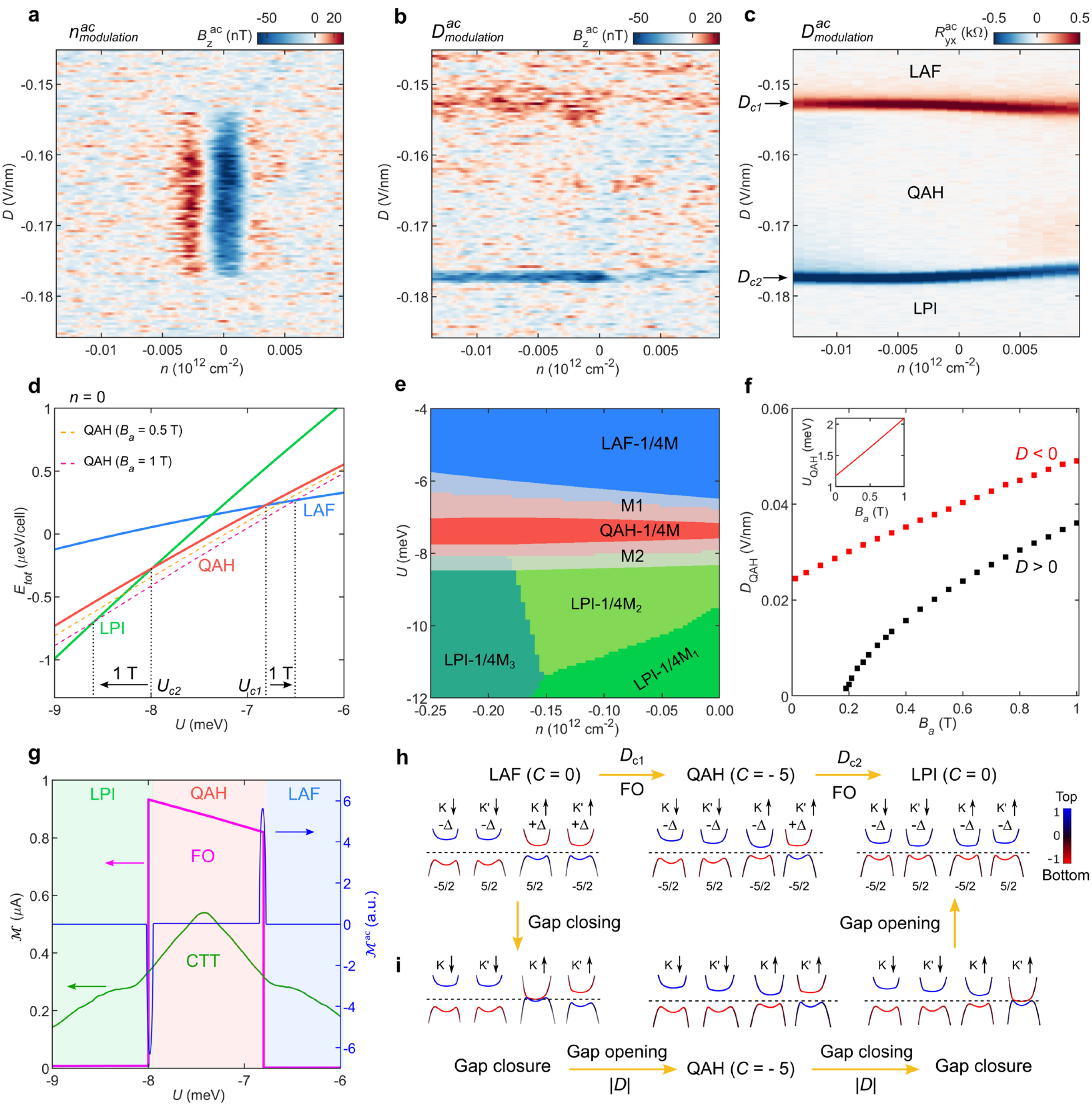

**Fig. 4. First-order QAH phase transition. a,** $B_z^{ac}(n, D)$ at a fixed location acquired at $B_a = 10.8$ mT and $T = 20$ mK using purely $n^{ac}$ modulation. **b,** The same as (a) but using purely $D^{ac}$ modulation. **c,** $D^{ac}$ modulated differential Hall resistance $R_{yx}^{ac} = D^{ac}\partial R_{yx}/\partial D$ as a function of $n$ and $D$ at $B_a = 12$ mT. Two sharp opposite polarity signals at $D_{c1} = -0.154$ V/nm and $D_{c2} = -0.178$ V/nm (arrows) indicate the FO phase transitions between the LAF, QAH, and LPI states, consistent with the local $B_z^{ac}$ response in (b) and in Extended Data Fig. 4. **d,** Calculated scHF energy $E_{tot}$ vs. displacement potential $U$ of the competing states at $n$ = 0: LAF (blue), LPI (green), and QAH at $B_a = 0$ T (red), 0.5 T (orange dashed), and 1 T (pink dashed). Black dotted lines indicate the QAH FO transitions at $B_a = 0$ T and 1 T (black arrows). **e,** Calculated scHF phase diagram of hole-doped R5G, featuring the 1/4M states of doped LAF and QAH, and three 1/4M variants of doped LPI. The shaded M1 and M2 areas show the expected fluctuation regions with minimal energy differences between neighboring states. **f,** Measured $B_a$ dependence of the width in the displacement field, $D_{QAH} = D_{c1} - D_{c2}$, of the QAH state at $D < 0$ (red) and the CI state at $D > 0$ (black). Inset: calculated $B_a$ dependence of $U_{QAH} = U_{c1} - U_{c2}$ for the QAH state as illustrated in (d). **g,** Calculated orbital magnetization $\mathcal{M}(U)$ as a function of displacement potential $U$ for first-order (FO, magenta) and continuous topological transition (CTT, green) at $B_a = 0$ T. Right axis: calculated differential magnetization

$\mathcal{M}^{ac} \cong d\mathcal{M}/dU$ for the FO transition (Methods). **h,** Illustration of FO LAF–QAH and QAH–LPI transitions as a function of negative $D$ (Supplementary Video 1), with the calculated scHF quasi-particle band structures of $K\uparrow$, $K'\uparrow$, $K\downarrow$, and $K'\downarrow$ flavors for $U = -5.2$ meV (LAF), $-7.6$ meV (QAH), and $-8.96$ meV (LPI), colored by layer polarization (red – bottom layer, blue – top layer). The Chern numbers $C$ and the signs of the individual gaps $\Delta$ are indicated. **i,** Schematics of continuous LAF–QAH and QAH–LPI transitions with smooth gap closure and reopening (Supplementary Video 2).

The band inversion across the FO transition produces a discontinuous change in the Berry curvature and, consequently, in the orbital magnetization and anomalous Hall conductivity. As a result, signatures of the transition are expected to remain observable even at finite $n$, when the Fermi level lies within the band edges rather than inside the QAH gap. To resolve the corresponding transport signatures, we apply a small $D^{ac}$, analogous to the local magnetization measurements, and measure the differential Hall response $R_{yx}^{ac} = D^{ac}\,\partial R_{yx}/\partial D$ . Figure 4c shows two sharp lines of peaks in $R_{yx}^{ac}$ that coincide with the corresponding features in $B_z^{ac}$ (Fig. 4b and Extended Data Fig. 4c), reflecting abrupt steps in $R_{yx}(D)$. Notably, the $B_z^{ac}$ and $R_{yx}^{ac}$ peaks (Figs. 4b,c and 2D $B_z^{ac}(x, y)$ images in Extended Data Fig. 3) persist to high hole doping, where the Fermi level lies well outside the QAH gap, demonstrating that the FO transitions extend deep into the metallic regime and that the discontinuous band inversion has a global impact on the orbital magnetic moment far beyond the gap region. Extended Data Figs. 4c,d further show that the measured peak widths in $B_z^{ac}$ and $R_{yx}^{ac}$ are determined by the amplitude of $D^{ac}$, fully consistent with a FO step in $\mathcal{M}$ and $R_{yx}$. Additional support for the FO nature of the transition is provided by hysteresis measurements of $R_{xx}$ and $R_{yx}$ upon sweeping $D$ across the transition (Extended Data Fig. 6). Although no resolvable hysteresis is observed at base temperature, a pronounced hysteresis develops upon warming, reaches its maximum near $T \approx 1.5$ K, and is strongest near charge neutrality, consistent with the largest magnetization discontinuity at the center of the QAH state.

We next quantify the stability range of the QAH state in $D$ , defining its width as $D_{\text{QAH}} = D_{c1} - D_{c2}$. Tracking the two $R_{yx}^{ac}$ peaks (Extended Data Fig. 4a) shows that $D_{\text{QAH}}$ grows approximately linearly with $B_a$ (Fig. 4f), indicating enhanced stabilization of the QAH state by $B_a$. The scHF calculations reproduce this behavior: the $E_{tot}$ of the QAH state decreases with $B_a$ due to the orbital contribution $-B_a \cdot \mathcal{M}$ in the Gibbs free energy ($B_a = 0.5$, 1 T, dashed orange/pink lines in Fig. 4d). Consequently, $U_{\text{QAH}} = U_{c1} - U_{c2}$, grows linearly with $B_a$ (Fig. 4f inset) and nearly doubles at 1 T (black dotted lines/arrows in Fig. 4d), closely reproducing the experimental observation. Importantly, this linear expansion constitutes additional strong confirmation of the FO QAH transition: a jump in $\mathcal{M}$ at the transition yields unequal shifts in free energy for the competing phases, expanding the QAH boundaries with $B_a$. By contrast, in a continuous transition the $\mathcal{M}$ of the competing phases at the transition should be equal (Fig. 4g, green curve), resulting in the same energy gain for the neighboring phases at the transition, and the boundary should thus be $B_a$ independent. Taken together, the discontinuous evolution of $\mathcal{M}$, the persistence of orbital-magnetic switching deep into the metallic regime, magnetic field expansion of the QAH phase, and the associated hysteretic behavior establish a direct thermodynamic signature of FO topological quantum phase transitions between the QAH state and its competing correlated states.

Figure 4e shows the phase diagram derived from our scHF calculations with the QAH state emerging as the ground state at intermediate negative displacement potentials $U$. Although quantized resistance occurs only at charge neutrality—where the Fermi level is in the quasiparticle band gap—the underlying band topology, characterized by three negative gaps and one positive gap (Fig. 4h), persists over a broad range of doping forming a new type of a quarter metal QAH-1/4M that is distinct from the previously reported 1/4M states [17–19] that arise from LPI phase (Extended Data Fig. 7 and Supplementary Video 3). The scHF calculations show that the range of QAH region (red in Fig. 4e) is only weakly dependent on hole doping, separated from neighboring phases by FO transitions. At small $|U|$, the LAF or LAF-1/4M state is the ground state, whereas at large $|U|$ and finite hole doping, several types of LPI-1/4M states become energetically favorable (Extended Data Fig. 7). LPI–1/4M$_1$ and LPI–1/4M$_2$ are fully isospin-polarized states carrying both spin and orbital magnetization. They are distinguished due to the Ising SOC, by whether the hole doping occurs in the $K\uparrow$ quasiparticle band, or instead in the $K'\uparrow$ band. In contrast, LPI–1/4M$_3$ is spin-polarized, valley-coherent 1/4M with no orbital magnetization. Consequently, the stability ranges of LPI–1/4M$_1$ and LPI–1/4M$_2$ are expected to grow with $B_a$, consistent with transport and local magnetization data (Extended Data Fig. 1).

This understanding sheds light on the contrasting behavior at positive and negative $D$. While a robust QAH state exists for $D < 0$ already at $B_a = 0$, a Chern insulator (CI) state at positive $D$ develops only above $B_a \cong$ 0.2 T, as shown by the black points in Fig. 4f (Extended Data Figs. 4b and 8b,c). This shows that for $D > 0$ the QAH phase has a higher $E_{tot}$ than the LAF and LPI phases, naturally explaining also the absence of fluctuations in the $D > 0$ metallic-like region, as discussed below. A finite $B_a$, however, lowers the energy of the magnetic QAH phase, similar to the behavior shown in Fig. 4d. The phase transition between this field-stabilized CI and the competing correlated phases is likewise FO, as evidenced by the sharp features in Extended Data Figs. 4b and 8e,f. Interestingly, in contrast to the QAH state at $D < 0$, the stability range of which is essentially $n$ independent (Figs. 4b,c,e), the CI state at $D > 0$ is rapidly suppressed with either electron or hole doping (Extended Data Fig. 8f). This indicates that at $D > 0$ the QAH phase is less stable against competing phases in the entire parameter space spanned by $D$, $B_a$, and $n$. This is likely because the electron-electron interaction, essential for the emergence of QAH phase, is weakened due to the $WS_2$ screening effect when the valence-band electrons are polarized to the top graphene layer ($D > 0$). Nevertheless, this calls for future investigation.

**Fluctuating strongly correlated phases**

We find that the metallic region M2 between QAH and LPI phases is characterized by strong fluctuations in transport measurements. The $R_{xx}$ (Fig. 5a) exhibits pronounced fluctuations reaching several kOhm—exceeding the average $R_{xx}$ value—across a broad range of both electron and hole doping, as well as over a wide span of $D$. Comparable fluctuations in both magnitude and extent are also observed in $R_{yx}$ (Fig. 5b). Here we present the first real-space imaging of the fluctuating phases revealing their spatial and temporal dynamics. Figure 5c presents the magnetic signal $B_z^{ac}$ measured at a single location in the sample vs. a narrow range in $n$ and $D$, marked by the dashed rectangle in Fig. 5a. The signal exhibits irregular variations with both positive and negative $B_z^{ac}$, consistent with fluctuating magnetic domains. Figures 5d-i show 2D $B_z^{ac}(x, y)$ images of the sample area at fixed $n$ and $D$ (magenta star in Fig. 5a,c), acquired sequentially in time over about 14 hours. The time sequence captures spontaneous emergence and disappearance of magnetic domains at different locations across the sample. This behavior persists over a range of doping on either side of charge neutrality, as shown by the sequence of images in Figs. 5j-o recorded at a fixed $D$ and a span of $n$ along the green arrow in Fig. 5c. Similar dynamic fluctuations are observed at a fixed $n$ and a span of $D$ as shown in Extended Data Fig. 9.

Extended Data Fig. 10 shows numerically reconstructed magnetization maps $\mathcal{M}$ over a wide range of $n$ at fixed $D$, revealing magnetic domains with characteristic sizes of ~ 1-2 μm coexisting with nonmagnetic regions. Strikingly, the magnitude of the reconstructed magnetization closely matches that of the QAH state (Figs. 2,3 and Extended Data Figs. 2,3). Since the QAH state is the only phase with substantial orbital magnetization among the competing correlated states, these magnetic domains must originate from local fluctuations of QAH regions with either positive or negative chirality embedded within the surrounding competing correlated phase. These observations indicate that the M2 regime is governed by coexistence and fluctuations of competing correlated phases, providing a microscopic manifestation of the first-order QAH transition.

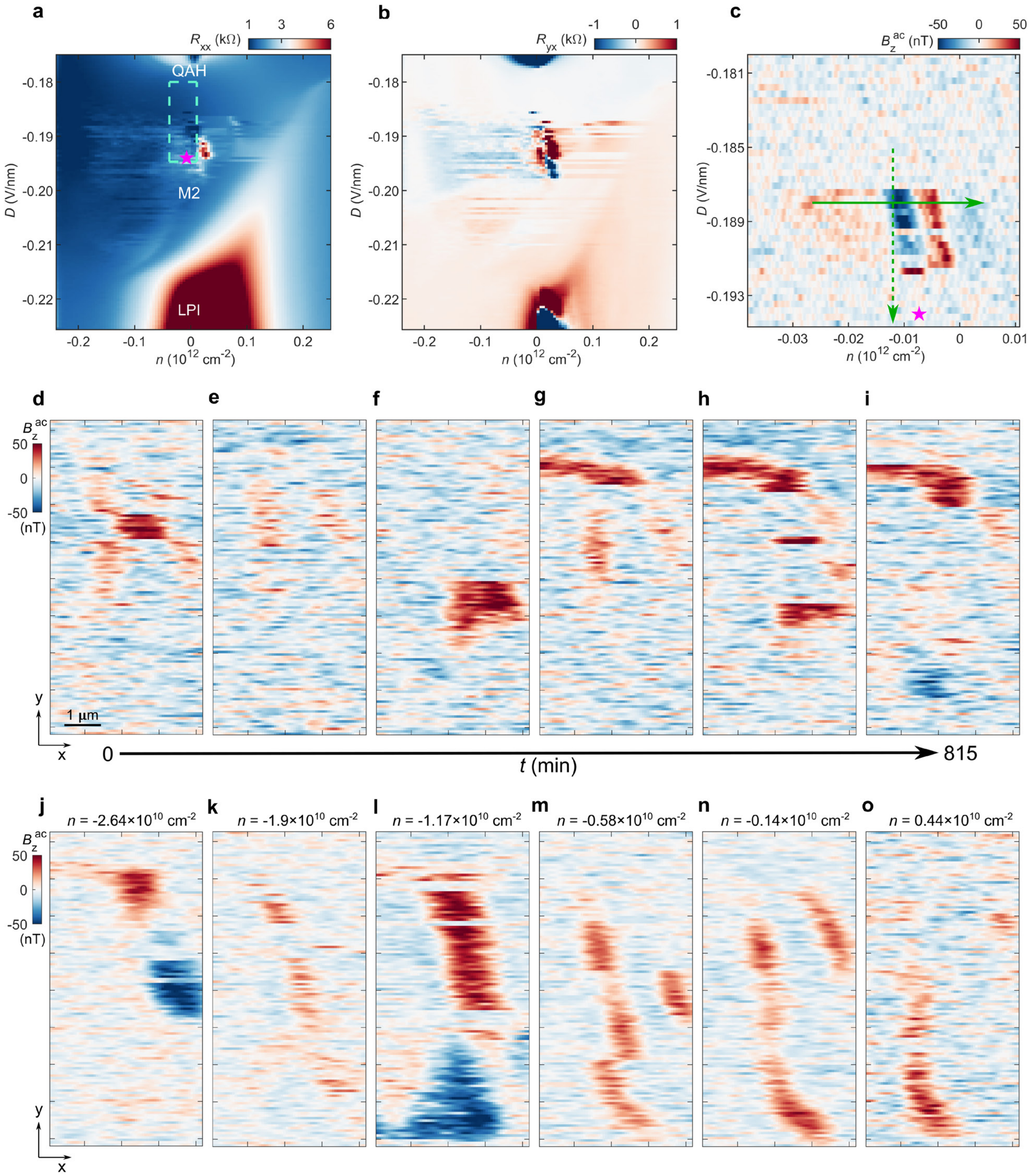

**Fig. 5. Fluctuating strongly correlated phases. a,** $R_{xx}(n, D)$ map in the vicinity of M2 state showing pronounced fluctuations. **b,** The same as (a) but for $R_{yx}$. **c,** $B_z^{ac}(n, D)$ measured in the area marked by a dashed frame in (a). **d-i,** Time-sequence of $B_z^{ac}(x, y)$ images of the sample area using $V_{bg}^{ac} = 6$ mV rms at fixed $n = -7.3\times10^9$ cm$^{-2}$ and $D = -0.194$ V/nm marked by the magenta stars in (a) and (c). The images were acquired consecutively with acquisition time of 163 mins per image from bottom-left to top-right, showing random fluctuations in time with abrupt jumps during the scan. **j-o,** $B_z^{ac}(x, y)$ images using $V_{bg}^{ac} = 6$ mV rms at various $n$ at $D = -0.188$ V/nm along the solid green arrow in (c), showing fluctuations over a wide range of doping. See Extended Data Fig. 9 for similar fluctuations along the dashed green arrow in (c). All measurements were performed at $B_a = 12$ mT and $T = 20$ mK.

Note that our measurements are sensitive only to fluctuations within a limited window of temporal and spatial scales, set by the integration time (~1 s) and spatial resolution (~100 nm). As a result, very rapid fluctuations or domains smaller than our spatial resolution may evade detection in both transport and magnetic imaging. Since the characteristic timescales and domain sizes of the fluctuations may span a broad range, it is likely that additional regions—appearing featureless in our measurements—are governed by similar dynamic mechanism, but with fluctuations occurring outside our accessible window.

**Discussion**

The QAH state realized in spin–orbit-proximitized rhombohedral graphene represents a distinct class of correlated topological matter. Unlike previously studied QAH systems [49–52], it is intrinsically intertwined with layer polarization, such that the Chern number, orbital magnetization, and electric polarization are all governed by the degree of layer polarization [5]. This unique connection enables direct control of Chern topology and orbital magnetization through the displacement field and gives rise to a rich landscape of competing ordered states. The local magnetic imaging presented here provides a microscopic view of these phases, their orbital magnetization, and the transitions between them.

The fluctuating M2 region provides a direct manifestation of this FO character. Local magnetic imaging reveals that this regime is not a homogeneous metallic phase but instead reflects coexistence and fluctuations between nearly degenerate states. The scHF calculations provide a natural framework for understanding this behavior. As shown in Fig. 4d, although the energies of the competing phases vary substantially with displacement field, the energy difference between the lowest-energy states remains small over a broad range surrounding the FO transitions. Such near degeneracy enhances the susceptibility to disorder, fluctuations, and phase coexistence. Consistently, the phase diagram in Fig. 4e identifies the M2 regime as a region where multiple states become nearly degenerate. In particular, the competition between QAH-derived and LPI-derived phases, together with the presence of several nearly degenerate LPI-related configurations, naturally promotes an extended fluctuation regime. By contrast, competing nearly degenerate states also exist near M1, but the manifold is smaller than that near M2. Consequently, the entropic stabilization of phase competition and coexistence is weaker near M1, consistent with the much weaker fluctuation behavior observed experimentally.

While the scHF calculations predict that the competing phases remain gapped across the FO transition, mean-field approaches are known to overestimate gap sizes. The microscopic evolution of the excitation spectrum across the fluctuating regime therefore remains an important question for future experimental and theoretical studies. Nevertheless, the experimental observations consistently point to a transition regime governed by strong competition between nearly degenerate phases, reflecting the first-order nature of the transitions and the resulting entropic fluctuations.

Finally, our observations establish a microscopic framework for understanding topological quantum phase transitions in strongly interacting quantum materials. In rhombohedral graphene, the transitions between topologically distinct phases undergo an extended regime of competition and fluctuations rather than through a simple homogeneous transformation. The intimate connection between layer polarization, topology, and orbital magnetization revealed here suggests that the degree of layer polarization acts as a unifying organizing principle across the correlated phase diagram at charge neutrality. This provides a new microscopic insight into recently observed switchable ferroelectric orbital magnetism in the semimetallic regimes near $D = 0$ across the rhombohedral multilayer graphene family [21,30–33], suggesting a likely origin in competing correlated states associated with different layer-polarization textures. More generally, our work demonstrates that local orbital magnetometry provides a powerful thermodynamic probe for revealing the interplay between quantum geometry, electronic ordering, and phase competition in strongly interacting quantum materials.

**Acknowledgments** We thank Erez Berg, Yuval Oreg, Yaar Vituri, Peleg Emanuel and Tobias Holder for fruitful discussions. This work was supported by the Minerva Stiftung with funding from the Federal German Ministry for Education and Research, by the United States-Israel Binational Science Foundation (BSF) grant No 2022013, and by the European Union (ERC, MoireMultiProbe - 101089714). Views and opinions expressed are however those of the author(s) only and do not necessarily reflect those of the European Union or the European Research Council. Neither the European Union nor the granting authority can be held responsible for them. E.Z. acknowledges the support of Tom and Mary Beck Center for Advanced and Intelligent Materials, Goldfield Family Charitable Trust, and Leona M. and Harry B. Helmsley Charitable Trust grant #2112-04911. E.Z. and L.J. acknowledge the MIT-Israel Zuckerman STEM Fund grant 2654504. L.J. acknowledges support from Nano & Material Technology Development Program through the National Research Foundation of Korea (NRF) funded by the Ministry of Science and ICT (RS-2024-004447252) and the MIT Portugal Program. T.H. acknowledges support from a Mathworks Fellowship. C.Y. and F.Z. acknowledge support from the National Science Foundation (NSF) under grants DMR-2414726, DMR-1945351, and DMR-2324033, support from the Welch Foundation under grant AT-2264-20250403, and supercomputing resources provided by the Texas Advanced Computing Center (TACC). K.W. and T.T. acknowledge support from the JSPS KAKENHI (Grant Numbers 20H00354, 21H05233 and 23H02052) and World Premier International Research Center Initiative (WPI), MEXT, Japan.

**Author contributions** N.A. designed and built the scanning SOT microscope. S.D. and N.A. performed the local magnetization measurements and S.D. analyzed the data. T.H. and Z.L. fabricated and characterized the samples. S.D. and M.U. performed the transport measurements and data analysis. Y.Z. and S.D. performed the magnetization reconstruction using the UPINN model. E.Z., S.D. and N.A. designed the experiment. C.Y. and F.Z. performed the scHF and phase diagram calculations. S.D. and C.Y. performed single-particle calculations. S.D., Y.M., N.S.K fabricated the SOT and the tuning fork, and M.E.H. developed the SOT readout. K.W. and T.T. provided the hBN crystals. S.D., C.Y., F.Z., L.J., and E.Z. wrote the original manuscript. All authors participated in discussions and revisions of the manuscript.

**Competing interests** The authors declare no competing interests.

**Data availability** The data that supports the findings of this study are available from the corresponding authors on reasonable request.

**Code availability** The scHF and single-particle calculations used in this study are available from the corresponding authors on reasonable request.

## Methods

### Device fabrication

The pentalayer graphene, monolayer $WS_2$ and hBN flakes were prepared by mechanical exfoliation onto $SiO_2$/Si substrates. The rhombohedral domains of pentalayer graphene were identified using near-field infrared microscopy [53], confirmed with Raman spectroscopy and isolated by cutting with a Bruker atomic force microscope (AFM) [54]. The van der Waals heterostructure was made following a dry transfer procedure. We picked up the top hBN, graphite, middle hBN, the $WS_2$, and the pentalayer graphene using polypropylene carbonate (PPC) film and landed it on a prepared bottom stack consisting of an hBN and graphite bottom gate. We intentionally misaligned the straight edges of the graphene and hBN flakes to avoid the effects of the moiré superlattices. The device was then etched into a Hall bar structure using standard e-beam lithography (EBL, Elionix F125) and reactive-ion etching (RIE). We deposited Cr/Au for electrical connections to the source, drain and gate electrodes.

### Nanoscale SOT measurements

Nanoscale magnetic imaging was performed using a home-built scanning SQUID-on-tip (SOT) microscope integrated into a dilution refrigerator operating at temperatures down to $T = 20$ mK [18]. For measurement, a indium SOT was used and fabricated on the apex of a sharp pipette following previously reported methods [45,55,56]. The SOT with diameter of $\sim$200 nm operated in magnetic fields of up to 0.3 T with sensitivity of $\sim$10 nT/Hz$^{1/2}$. To maintain a constant scan height, the SOT was attached with a quartz tuning fork force sensor [57] oscillating at its resonance frequency of $\sim$32 kHz. The SOT signal was read out using a cryogenic series SQUID array amplifier [58]. A small *ac* voltages $V_{tg}^{ac}$ and $V_{bg}^{ac}$ at frequency $f \approx 1.32$ kHz were applied to top and back gates in addition to the *dc* voltages $V_{tg}^{dc}$ and $V_{bg}^{dc}$ to modulate the carrier density $n^{ac} = (C_{tg}V_{tg}^{ac} + C_{bg}V_{bg}^{ac})/e$ and displacement field $D^{ac} = (C_{tg}V_{tg}^{ac} - C_{bg}V_{bg}^{ac})/2\varepsilon_0$. For pure $n^{ac}$ modulation measurement (Fig. 4a), we applied $V_{tg}^{ac} = 2.25$ mV rms and $V_{bg}^{ac} = 1.5$ mV rms. For pure $D^{ac}$ modulation measurements (Fig. 4b,c), we used $V_{tg}^{ac} = -18$ mV rms and $V_{bg}^{ac} = 12$ mV rms. All the 2D spatial images were recorded by imaging at constant height $\sim$200 nm above the sample surface with pixel size $\sim$55 nm.

### Transport measurements

The transport data were acquired using standard low frequency lock-in measurement technique at the base temperature of the dilution refrigerator of 20 mK, using excitation current of 1–5 nA at $f \approx 11.3$ Hz. For $D^{ac}$ modulated differential resistance a small *ac* voltages (same amplitude and frequency as used in the local SOT measurements) were superimposed on the *dc* voltages $V_{tg}^{dc}$ and $V_{bg}^{dc}$ and a constant *dc* current $I_{dc} = 5$ nA, applied through 1 Mohm bias resistor. The corresponding differential resistances were defined as, $R_{xx}^{ac} = \frac{V_{xx}^{ac}}{I_{dc}}$ and $R_{yx}^{ac} = \frac{V_{yx}^{ac}}{I_{dc}}$, respectively, where $V_{xx}^{ac}$ and $V_{yx}^{ac}$ are the longitudinal and transverse voltages measured at the modulation frequency.

### Symmetry-broken metallic states

We probed the isospin degeneracy of metallic phases adjacent to the charge-neutral correlated insulators by measuring $R_{xx}$ under $B_a = 3$ T. The flavor degeneracy $\nu$ was derived from the quantum oscillation (QO) period in carrier density, $\Delta n = \frac{\nu e B_a}{h}$. At high hole doping, the QOs show a series of Stoner-type symmetry-breaking transition from full metal (M) to partially isospin polarized (PIP) metal to half metal (s-1/2M), as reported previously [17–20]. At lower density, the phase map contains primarily simple (s-1/4M) and annular (a-1/4M) quarter metal phases (Extended Data Fig. 1c), consistent with the qualitative trends in the scHF phase map (Fig. 4e and Extended Data Fig. 7). The boundary of a-1/4M phase shows a sharp change in both $R_{xx}$ and $B_z^{ac}$ (Extended Data Figs. 1a,d). With increasing $B_a$, the a-1/4M phase expands towards lower $|D|$ for $D < 0$ (Extended Data Figs. 1b,e,f), in contrast to bare R4LG where the 1/4M phases grow towards higher $D$ [18]. This expansion can be understood qualitatively within our scHF theory. In this $D$ range, the scHF calculations reveal three distinct 1/4M phases emerging from the LPI state with hole doping: LPI-1/4M$_3$ (valley-coherent, without orbital magnetization) and LPI-1/4M$_1$ and LPI-1/4M$_2$ (both spin and valley polarized with finite orbital magnetization). Consequently, LPI-1/4M$_1$ and LPI-1/4M$_2$ regions expand with $B_a$, suggesting that the observed a-1/4M originates from these LPI-derived 1/4M phases.

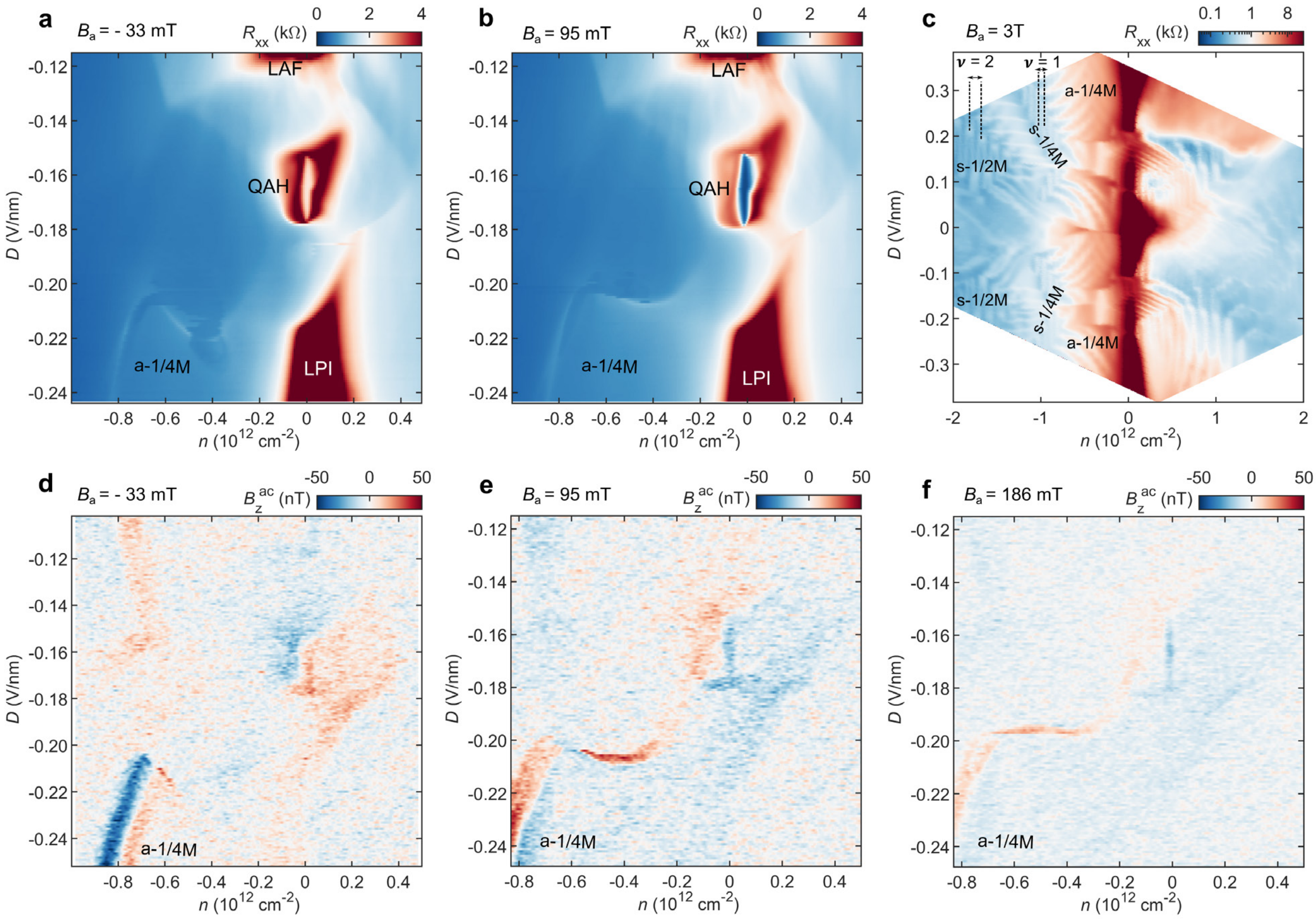

**Extended Data Fig. 1. Symmetry-broken metallic states. a,** $R_{xx}\ (n, D)$ measured in the vicinity of QAH state at $B_a = -33$ mT. **b,** Same as (a) but at $B_a = 95$ mT. **c,** $R_{xx}\ (n, D)$ measured at $B_a = 3$ T over a wide range of $n$ and $D$. **d,** $B_z^{ac}(n, D)$ measured at $B_a = -33$ mT with $V_{bg}^{ac} = 80$ mV rms. **e,** Same as (d) but at $B_a = 95$ with $V_{bg}^{ac} = 60$ mV rms. **f,** Same as (d) but at $B_a = 186$ mT with $V_{bg}^{ac} = 40$ mV rms, showing expansion of the a-1/4M phase with $B_a$.

### Orbital magnetization patterns in the gapped QAH state and across the FO transition in metallic states

Extended Data Fig. 2 shows the evolution of orbital magnetization with $D$ at a fixed small nominal $n$ close to charge neutrality where most of the sample is in the gapped QAH state. The $B_z^{ac}(x, y)$ patterns show little dependence on $D$ indicating that the inhomogeneity arises mainly due to charge disorder in the sample.

The QAH FO transition has a global effect on orbital magnetization and Hall resistivity, resulting in sharp peaks in $B_z^{ac}$ and $R_{yx}^{ac}$ across the transitions even in the metallic states far away from charge neutrality as shown in Figs. 4b,c. Extended Data Figs. 3a-j show the spatial maps of $B_z^{ac}(x, y)$ at the FO transition between metallic QAH-1/4M and LPI-1/4M states along $D_{c2}$ line over a significantly wider range of carrier densities $n$ than in Fig. 4b. Note that the origins of the $B_z^{ac}$ signals in Extended Data Figs. 2 and 3 are distinct. In Extended Data Fig. 2, the negative $B_z^{ac}$ reflects the negative differential magnetization in the gapped QAH state that arises due to modulation in the Chern magnetization, $\partial\mathcal{M}/\partial\mu = C\frac{e}{h}$ with $C = -5$. In contrast, in Extended Data Fig. 3 the negative $B_z^{ac}$ reflects the discontinuous drop in $\mathcal{M}$ across the FO transition between metallic QAH-1/4M and LPI-1/4M states, which mainly arises from the difference in self-rotation magnetization $\mathcal{M}_{SR}$.

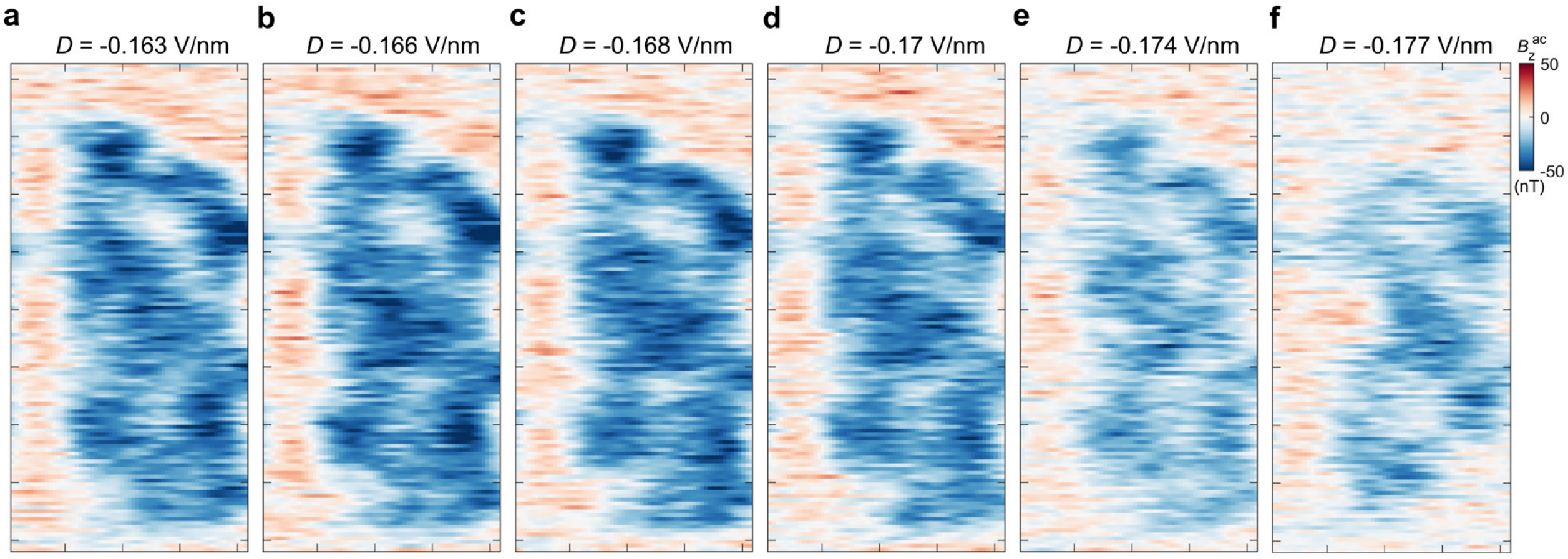

**Extended Data Fig. 2. Evolution of orbital magnetization with $D$ in the QAH state. a-f,** $B_z^{ac}(x, y)$ maps at $B_a = 12$ mT with $V_{bg}^{ac} = 6$ mV rms, measured close to charge neutrality at a fixed $n = -0.96 \times 10^9$ cm$^{-2}$ (in-between Figs. 2i and 2j) as a function of $D$ within the QAH state range $D_{c2} < D < D_{c1}$.

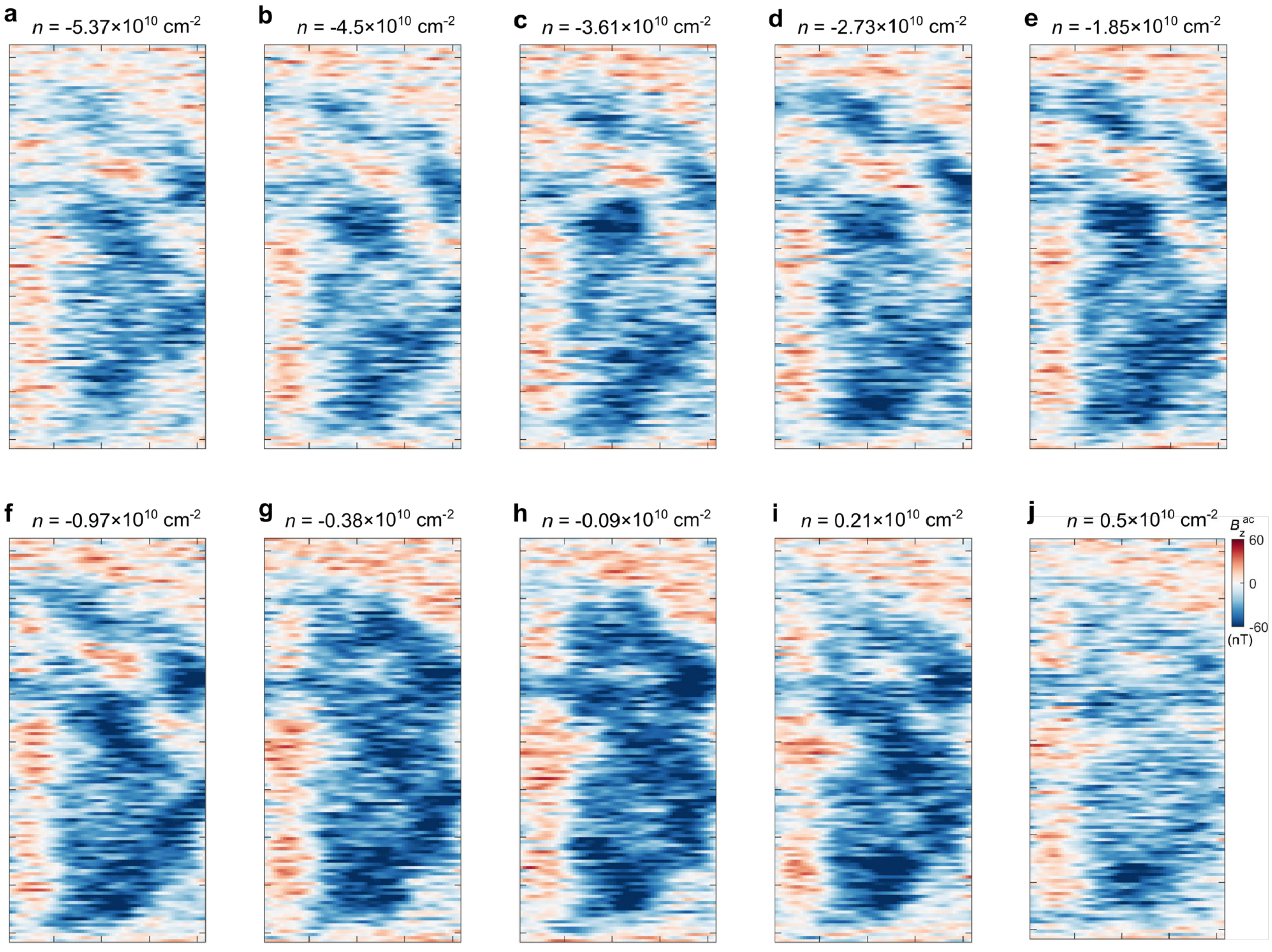

**Extended Data Fig. 3. Orbital magnetization at FO transition between metallic states. a-j,** Measured $B_z^{ac}(x, y)$ as a function of $n$ at FO transition between QAH-1/4M and LPI-1/4M at $D_{c2} = -0.1773$ V/nm and $B_a$ = 10.8 mT using $V_{bg}^{ac} = 12$ mV rms.

### Evolution of $B_z^{ac}$ and $R_{yx}^{ac}$ across FO and continuous transitions

To interpret the peaks observed in $B_z^{ac}$ and $R_{yx}^{ac}$ across the FO transition with $D^{ac}$ modulation (Extended Data Fig. 4), we computed the band structure and $\mathcal{M}$ as a function of $U$ using scHF theory that yields a FO transition. The scHF calculations of the evolution of the band structure, gap size, and $\mathcal{M}$ with $U$ are presented in Supplementary Video 1, Extended Data Fig. 5d, and Fig. 4g, respectively, showing FO transitions in all the quantities.

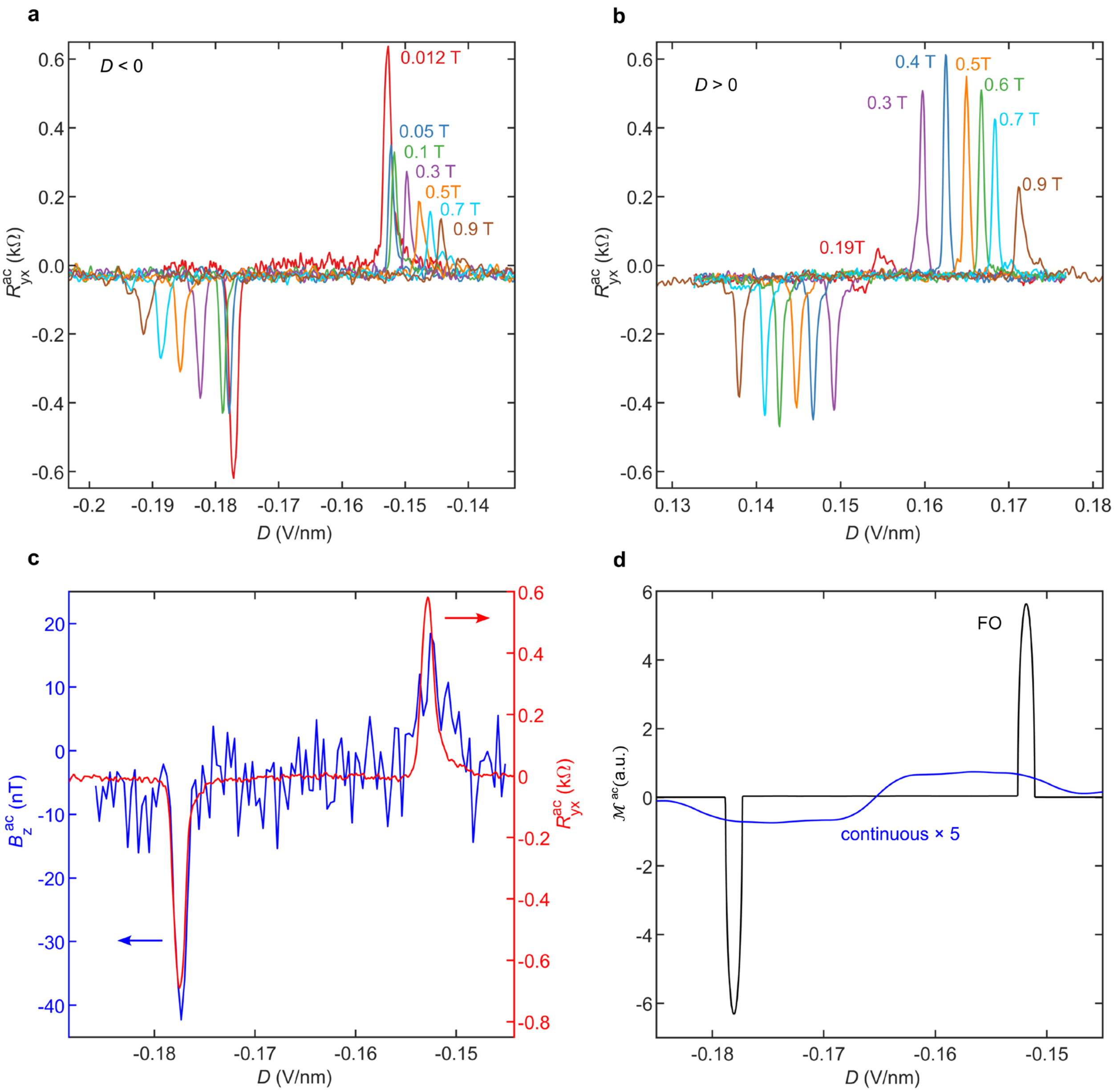

**Extended Data Fig. 4. Analysis of $B_z^{ac}$ and $R_{yx}^{ac}$ across FO and continuous transitions**. **a,** Line traces of $R_{yx}^{ac}$ vs. $D$ measured at various $B_a = 0.012$ (red), 0.05 (blue), 0.1 (green), 0.3 (purple), 0.5 (orange), 0.7 (cyan), and 0.9 T (brown) across the QAH state for $D < 0$. At each $B_a$ the $n$ was chosen to be in the center of the gapped state according to Streda formula. **b,** Same as (a) but for the CI state at $D > 0$ measured at $B_a = 0.19$ (red), 0.3 (purple), 0.4 (blue), 0.5 (orange), 0.6 (green), 0.7 (cyan), and 0.9 T (brown). The separation between the positive and negative peaks, $D_{\text{QAH}}$, vs. $B_a$ is plotted in Fig. 4f. **c,** Linecut of $B_z^{ac}$ (left) and $R_{yx}^{ac}$ (right) vs. $D$, obtained from Fig. 4b,c by averaging over five lines centered at $n = -1.3\times10^{10}$ cm$^{-2}$. **d,** Simulation of the experimental $\mathcal{M}^{ac}$ using theoretically calculated $\mathcal{M}(U)$ at $n = -1.0\times10^{10}$ cm$^{-2}$ using $D^{ac}$ modulation for the cases of FO (black) and continuous (blue, enlarged five times) transitions.

For comparison, we carried out complementary non-interacting ad-hoc calculations using a ten-band tight-binding model with gap signs assigned as $\{+,-,+,-\}$ for flavors $\{K\uparrow, K\downarrow, K'\uparrow, K'\downarrow\}$, mimicking the LAF state. To match the experimentally measured $D_{c1}$ and $D_{c2}$ end points of the QAH state we chose the intrinsic gap $\Delta_{int}$= 33 meV at $D = 0$ with an additional spin-orbit coupling term $\lambda_I = 4.91$ meV, and used the previously reported conversion of $D = 1$ V/nm corresponding to $U/2 = 100$ meV [18,46]. The resulting evolution of the band structure, the global gap size, and $\mathcal{M}$ with $D$ is presented in Supplementary Video 2 and Extended Data Fig. 5f, showing continuous behavior of all these properties. In this case, at almost any $D$ the different flavors have unequal gaps. As a result, at low doping the system is in an effective 1/4M state with finite $\mathcal{M}$ for all three LAF, QAH, and LPI states. The calculated $\mathcal{M}(D)$ (purple in Extended Data Fig. 5f) evolves smoothly and decays on both sides of the QAH state, inconsistent with the experimental findings.

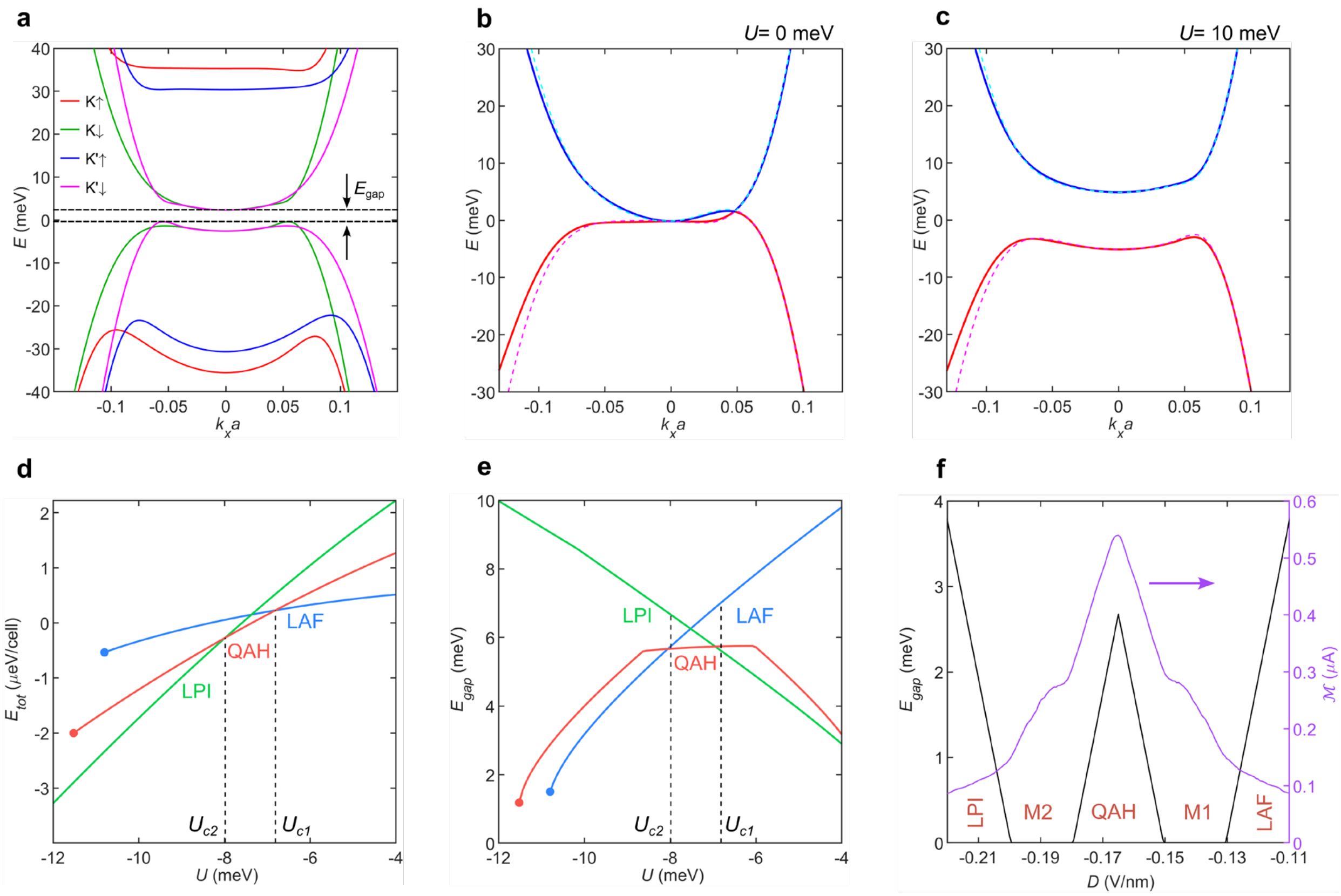

**Extended Data Fig. 5. Evolution of band parameters across FO and continuous topological transitions. a,** Low-energy band structures of the four flavors $\{K\uparrow, K\downarrow, K'\uparrow, K'\downarrow\}$ in the middle of the QAH state at $D = -0.165$ V/nm, computed using the phenomenological continuous topological transition model. (see Supplementary Video 2 for band structure evolution with $D$). The dashed lines define the global energy gaps $E_{gap}$. **b,** Low-energy dispersion in the $K$ valley for $U = 0$, comparing the ten-band tight-binding model with $\Delta_{int} = \lambda_I = 0$ (dashed lines) to the effective two-band model (solid lines) used in the scHF calculations. **c,** Same as (b) but for $U = 10$ meV. See Supplementary Video 1 for band structure evolution with $U$ calculated using scHF theory. **d,** Calculated scHF energy $E_{tot}$ vs. $U$ at $n = 0$ and $B_a = 0$ T of LAF, QAH, and LPI states as in Fig. 4d but over extended range of $U$. Black dashed lines indicate the QAH FO transitions at $U_{c2}$ and $U_{c1}$. Red and blue dots indicate the end points of the stability range of QAH and LAF states where their gaps close, as marked in (e). **e,** Calculated scHF global energy gaps $E_{gap}$ vs. $U$ of the three insulator states. At the FO transitions at $U_{c2}$ and $U_{c1}$ the gaps of all the states are finite and large. **f,** Global energy gap $E_{gap}$ (left axis) vs. $D$ calculated using the same phenomenological continuous topological transition model. The gap closes continuously at the boundaries of the QAH state leading to a smooth evolution of orbital magnetization $\mathcal{M}$ (right axis), inconsistent with the experiment.

Note that at finite doping, the finite $\mathcal{M}$ at the gap closure arises from the three bands that retain finite but distinct gaps at $D_{c1}$ and $D_{c2}$ (Supplementary Video 2), while the continuously closing and reopening gap of the remaining band contributes no discontinuity. Consequently, both $\mathcal{M}$ and $R_{yx}$ remain finite yet evolve smoothly across the transition, despite the sign reversal of one of the band gaps.

To model the experimental $D^{ac}$ modulation, at each $U$ we apply a small *ac* drive, $U(t) = U + U^{ac}\sin(\omega t)$, where $\omega$ is the modulation frequency with modulation amplitude $U^{ac} = \frac{U}{D}D^{ac}$. For each $U$, we compute $\mathcal{M}(t)$ and derive the first-harmonic $\mathcal{M}^{ac}$ of the fast Fourier transform of $\mathcal{M}(t)$, as effectively measured in the experiment. For the FO transition in scHF model, where $\mathcal{M}(U)$ jumps at $U_{c1}$and $U_{c2}$ from $\mathcal{M}_{QAH}$ to almost zero, the modulated response $\mathcal{M}^{ac}$ becomes a semicircular peak centered at $U_{c1}$ and $U_{c2}$ with peak height of $2\mathcal{M}_{QAH}/\pi$ and width of $2U^{ac}$. The derived peak width in Fig. 4g and Extended Data Fig. 4d agrees quantitatively with the experimental peak width in $R_{yx}^{ac}$ and $B_z^{ac}$ at the QAH-to-LPI transition in Extended Data Fig. 4c. In contrast, applying the same analysis to the continuous gap closing and reopening model, yields only weak and broad features in $\mathcal{M}^{ac}$ without any sharp peaks (blue curve in Extended Data Fig. 4d), providing additional confirmation of the FO LAF-QAH-LPI phase transitions.

**Hysteresis at the QAH transition**

Hysteresis is one of the hallmarks of FO phase transitions, reflecting metastability and the coexistence of competing phases separated by an energy barrier. To probe hysteresis across the QAH transitions, we measured $R_{xx}$ and $R_{yx}$ while sweeping $D$across the transition points at several carrier densities (dashed lines in Extended Data Fig. 6a) and at various temperatures. Although no resolvable hysteresis is observed at the base temperature, the hysteresis develops progressively with increasing temperature and is most pronounced at $T \approx 1.5$ K, as shown in Extended Data Figs. 6c–l. The hysteresis is strongest near charge neutrality (Extended Data Figs. 6f,g,k,l) and gradually diminishes with increasing electron or hole doping. The observation of hysteresis provides independent evidence supporting the FO nature of the transition. At higher temperatures, the hysteresis is suppressed as the transition becomes thermally broadened. In contrast, no measurable hysteresis is observed at the $D_{c1}$ transition over the entire temperature range studied, consistent with the substantially smaller magnetization discontinuity at $D_{c1}$ observed in Fig. 4b.

**Self-consistent Hartree-Fock (scHF) calculations**

In our scHF calculations, we adopt an effective two-band Hamiltonian (per spin-valley), derived from the full π-band tight-binding model [13,38] : $H_{\text{eff}} = \left[\sum_{n=0}^{4}\left(\frac{v_0 p}{\gamma_1}\right)^{2n}\right]^{-1} \times H_{\text{eff},0}$, where $H_{\text{eff},0}$ is given by

$$H_{\text{eff},0} = \frac{(v_0 p)^5}{\gamma_1^4}\left[\cos(5\phi_{\boldsymbol{k}})\,\sigma_x + \sin(5\phi_{\boldsymbol{k}})\,\sigma_y\right] + \left(\frac{3\gamma_2}{2} - \frac{4v_0 v_3 p^2}{\gamma_1}\right)\frac{(v_0 p)^2}{\gamma_1^2}\left[\cos(2\phi_{\boldsymbol{k}})\,\sigma_x + \sin(2\phi_{\boldsymbol{k}})\,\sigma_y\right]$$
$$+\left[\delta - \frac{2v_0 v_4 p^2}{\gamma_1}\sum_{n=0}^{3}\left(\frac{v_0 p}{\gamma_1}\right)^{2n}\right]\sigma_0 + \frac{U}{2}\left[\sum_{n=0}^{4}\left(1-\frac{n}{2}\right)\left(\frac{v_0 p}{\gamma_1}\right)^{2n}\right]\sigma_z + \lambda_I \tau_z s_z \frac{(\sigma_0 - \sigma_z)}{2},$$

with $\phi_{\boldsymbol{k}} = \tau_z \tan^{-1}\left(\frac{p_y}{p_x}\right)$, $\sigma_z = \pm 1$ denoting the 1A and 5B sublattices, and $\tau_z = \pm 1$ labeling the $K$ and $K'$ valleys. The parameter $U$ represents the electric potential difference between the two outermost graphene layers. Here, $U < 0$ corresponds to $D < 0$ in our experiment. $H_{\text{eff},0}$ captures the low-energy physics up to eighth order in $v_0 p$ and to first order in $\gamma_2$, $v_3 p$, and $v_4 p$ (with $v_i = \frac{\sqrt{3}a}{2\hbar}\gamma_i$) associated with remote hopping processes, relative to the nearest-neighbor interlayer hopping $\gamma_1$. We obtained the hopping parameter values from first-principles calculations: $\gamma_0 = 3160$, $\gamma_1 = 435$, $\gamma_2 = -18.5$, $\gamma_3 = -322$, $\gamma_4 = -67.5$, and $\delta = -0.147$ meV. The comparison with the full ten-band model confirms that the two-band model reliably reproduces the low-energy band structure, as shown in Extended Data Figs. 5b,c.

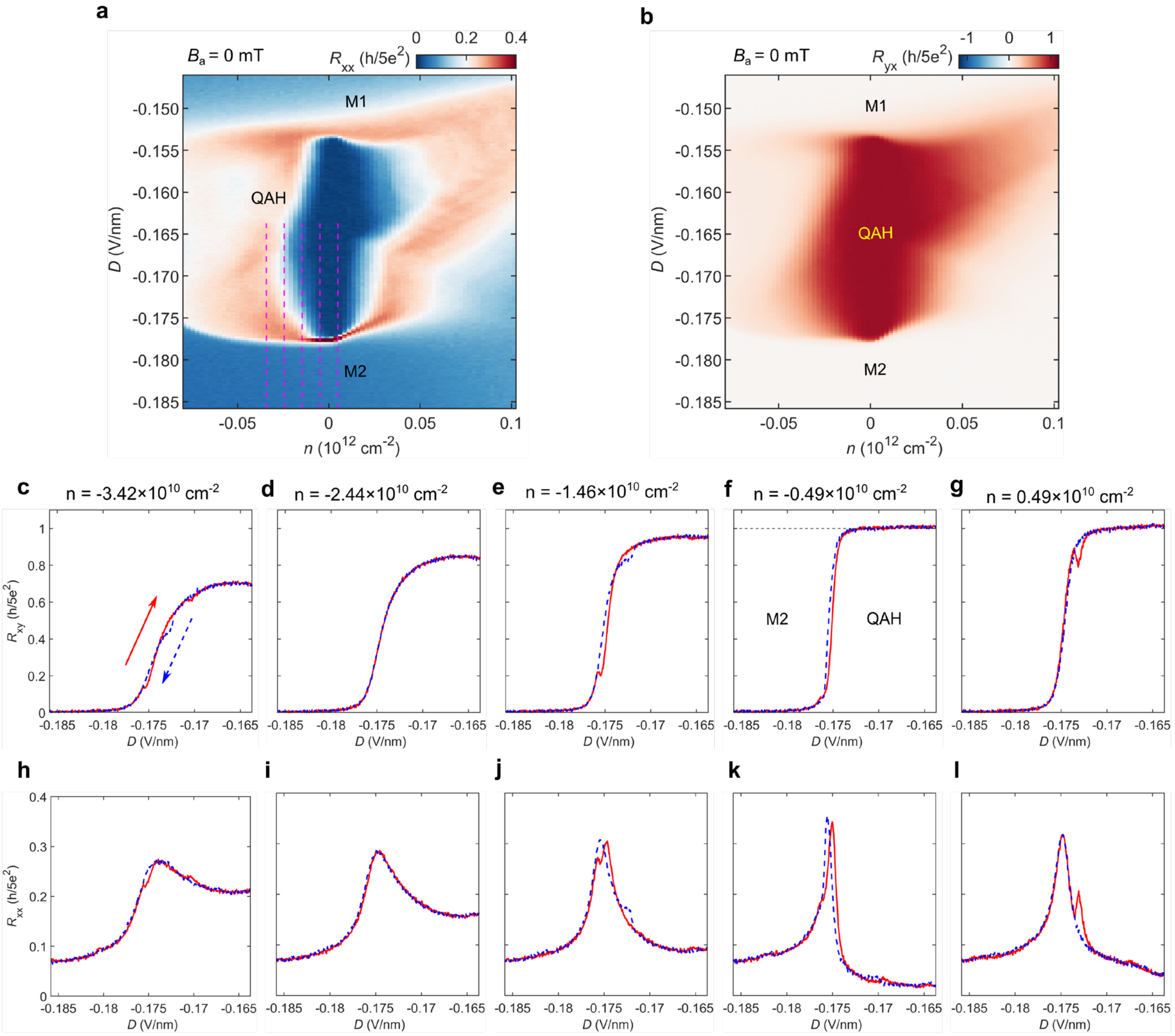

**Extended Data Fig. 6. Hysteresis at the QAH-M2 transition. a,** Longitudinal resistance $R_{xx}(n, D)$ in the vicinity of the QAH state measured at $T = 1.5$ K and $B_a = 0$ mT. **b,** Corresponding Hall resistance $R_{yx}(n, D)$. **c-g,** $R_{xy}$ as a function of $D$ at fixed $n$ indicated by the dotted magenta lines in (a), showing hysteresis across the QAH–M2 transition. **h-l,** Corresponding $R_{xx}(D)$ traces at the same carrier densities, showing hysteresis across the transition.

The term proportional to $\lambda_I$ originates from the Ising spin-orbit coupling (SOC) of the $WS_2$ layer adjacent to the top graphene layer [59], with a projector $\frac{(\sigma_0 - \sigma_z)}{2}$ reflecting this experimental configuration. In our calculations, we set $\lambda_I = 1$ meV in line with previous estimates [38,60,61]. The Rashba-type SOC is not included in our calculations, as its role has been shown to be negligible [38].

For calculating the band structure of R5G we use a $301 \times 301$ $k$-point square mesh grid centered at the $K/K'$ point with a side length of 0.3 in units of the inverse of graphene lattice constant. By comparing the total energies of competing interacting phases, we identify the ground state configuration. Figure 4h and Extended Data Fig. 7 show some representative scHF quasiparticle band structures.

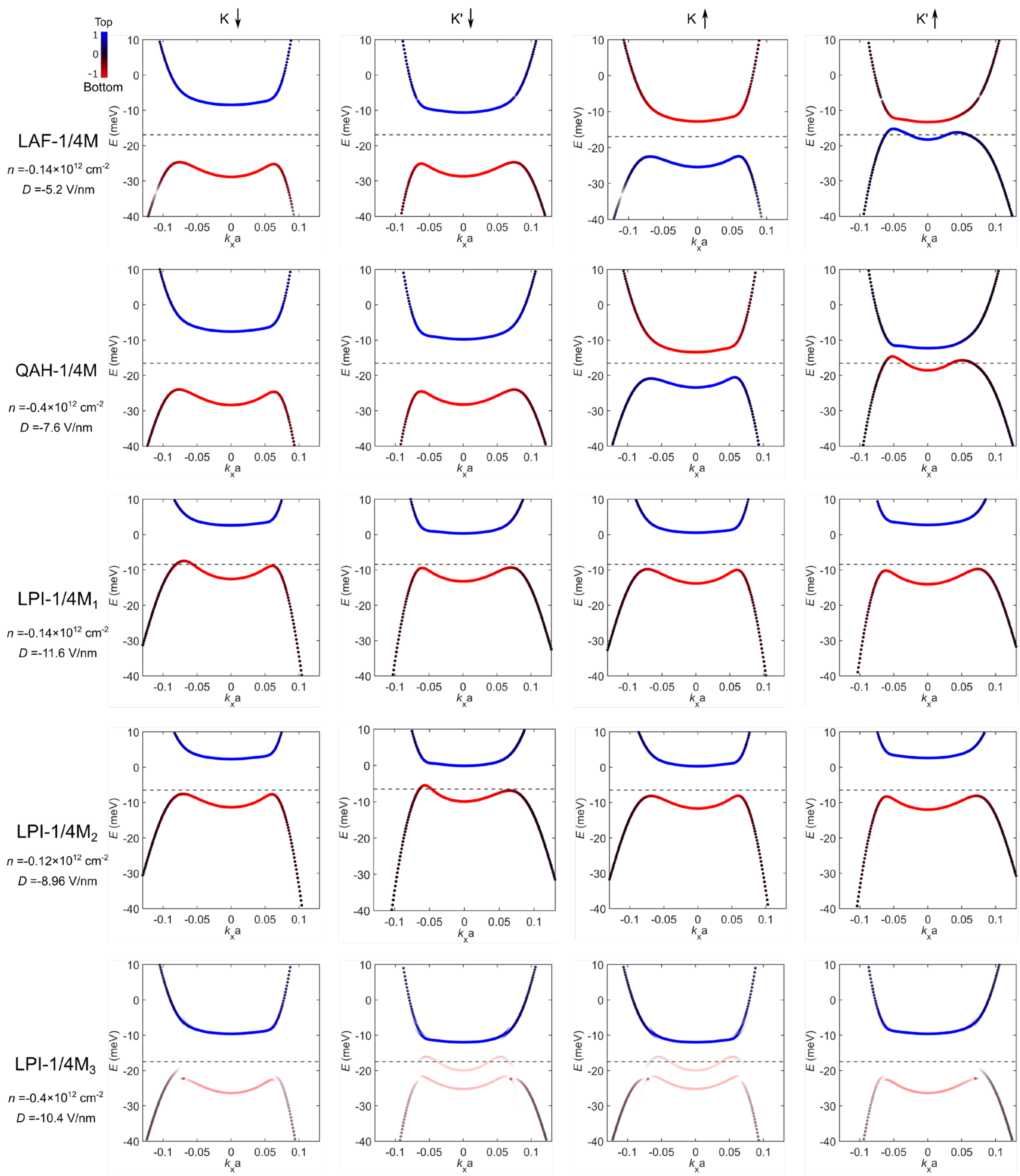

**Extended Data Fig. 7. Doping induced 1/4M phases from correlated insulators.** The quasiparticle band structures of the 1/4M phases for the four isospin flavors $\{K\uparrow, K'\uparrow, K\downarrow, K'\downarrow\}$ are obtained within the scHF theory at finite doping. Color intensity encodes layer polarization. The 1/4M phases emerge from the charge-neutral correlated insulators upon hole doping, without altering their initial layer polarization. See Supplementary Videos 1 and 3 for band structure evolution with $U$ and doping.

The exchange interaction, which involves momentum transfer within a single valley, is described by

$$\langle \hat{V}_{\mathrm{ex}} \rangle = -\frac{1}{2A} \sum_{\boldsymbol{k} s \tau \alpha} \sum_{\boldsymbol{k}' s' \tau' \alpha'} V_{\boldsymbol{k}-\boldsymbol{k}'}^{\alpha\alpha'} n_{\tau' s' \alpha', \tau s \alpha}(\boldsymbol{k}') n_{\tau s \alpha, \tau' s' \alpha'}(\boldsymbol{k}) \,,$$

where $A$ represents the system area, and $\boldsymbol{k}$ is the momentum relative to the $K/K'$ point. The indices $s, \tau$, and $\alpha$ denote the spin, valley, and orbital (1A, 5B) degrees of freedom, respectively. The density matrix is defined as $n_{\tau s\alpha,\tau' s'\alpha'}(\boldsymbol{k}) = \langle \hat{c}^{\dagger}_{\boldsymbol{k},\tau s\alpha} \hat{c}_{\boldsymbol{k},\tau' s'\alpha'} \rangle$, with $\hat{c}$ and $\hat{c}^{\dagger}$ being the annihilation and creation operators, respectively. The Coulomb potential $V_{\boldsymbol{q}}^{\alpha\alpha'}$, incorporating screening by dual metallic gates via the image charge method [62], takes the form:

$$V_{\boldsymbol{q}}^{\alpha\alpha'} = \begin{cases} \frac{2\pi e^2}{\epsilon_r q} \frac{\cosh(2qd_{\text{gate}}) - \cosh(2qd)}{\sinh(2qd_{\text{gate}})} & (\alpha = \alpha'), \\ \frac{2\pi e^2}{\epsilon_r q} \frac{\cosh[2q(d_{\text{gate}} - d)] - 1}{\sinh(2qd_{\text{gate}})} & (\alpha \neq \alpha'), \end{cases}$$

where $d = 3.35$ Å is the graphene interlayer spacing and $d_{\text{gate}}$ is the distance from the metallic gate to the center of R5G. We use $\epsilon_r = 35$ and $d_{\text{gate}} = 37$ nm to match the experimental result.

The Hartree interaction, capturing direct Coulomb repulsion, is given by

$$\langle \hat{V}_{\text{H}} \rangle = \frac{1}{2A} \sum_{\boldsymbol{k}} \sum_{s\tau\alpha} \sum_{s'\tau'\alpha'} V_{\boldsymbol{0}}^{\alpha\alpha'} n_{\tau' s'\alpha',\tau' s'\alpha'}(\boldsymbol{k}) n_{\tau s\alpha,\tau s\alpha}(\boldsymbol{k}).$$

This interaction screens the externally applied gate-displacement field. While the exchange interaction favors net layer polarization—lowering energy when isospins share similar wave functions—the Hartree interaction penalizes such polarization. The competition between Hartree and exchange interactions, combined with the gate-displacement field coupling to its layer polarization, gives rise to the emergence of the QAH phase studied in this work.

Finally, the valley-interchange (VI) exchange interaction [6,38], associated with large-momentum transfer across valleys, is

$$\langle \hat{V}_{\text{VI}} \rangle = -\frac{1}{A} \sum_{\boldsymbol{k}s\alpha} \sum_{\boldsymbol{k}' s'\alpha'} V'^{\alpha\alpha'} n_{\text{K}' s'\alpha',\text{K}' s\alpha}(\boldsymbol{k}') n_{\text{K}s\alpha,\text{K}s'\alpha'}(\boldsymbol{k}),$$

where a constant Coulomb potential $V'^{\alpha\alpha'} = \alpha_{\text{VI}} V_{\boldsymbol{0}}^{\alpha\alpha'}$ is used, as the transferred momentum is approximately the distance between the two valleys. In our calculations, $\alpha_{\text{VI}} = 0.03$ is chosen to match with the experimental data. Although the VI exchange energy is much smaller than the Hartree and exchange interactions, making it almost negligible for the determination of QAH phase boundary, it becomes important in two situations: (i) in the small-$|D|$ regime, where it selects the ground state among the 5 possible phases [5,6]; and (ii) in determining the ground state among the 1/4M states emerging from the hole-doped LPI state. The VI exchange interaction disfavors valley polarization, leading to the LAF ground state at charge neutrality and the spin-polarized, valley-coherent LPI-1/4$M_3$ state at high hole doping, respectively.

The VI exchange competes with the Ising SOC, which dictates antiparallel out-of-plane spin-valley locking near the band edge. As a result, dominant VI exchange yields in-plane spin quantization, whereas strong Ising SOC stabilizes a correlated quantum spin Hall insulator with out-of-plane spin quantization. When the two effects are comparable, spin canting emerges. Our calculations show that the out-of-plane canting angle of the LAF phase is as small as 9.6°, which we ignore in the main text for simplicity.

**Orbital magnetization and magnetic field dependence at $D < 0$**

In scHF, the orbital magnetization is evaluated using the following expression, which can be derived either from the semiclassical wave-packet formalism [40,42] or from the Wannier function theory [41]:

$$\mathcal{M} = -\frac{ie}{2\hbar} \sum_{n \in occ.} \int \frac{d\boldsymbol{k}}{(2\pi)^2} \langle \nabla_{\boldsymbol{k}} u_{n\boldsymbol{k}} | \times \left[ \hat{H}_{\boldsymbol{k}} + \varepsilon_{n\boldsymbol{k}} - 2\mu \right] | \nabla_{\boldsymbol{k}} u_{n\boldsymbol{k}} \rangle,$$

where $\varepsilon_{n\boldsymbol{k}}$ and $|u_{n\boldsymbol{k}}\rangle$ are the energy eigenvalue and the periodic part of the Bloch wave function for the $n$th occupied band, respectively. Note that the validity of the formula has been tested numerically and turns out to be accurate [41]. The expression for the total magnetization $\mathcal{M}$ can be regrouped into two terms [6,40–42,45]: self-rotation magnetization $\mathcal{M}_{\text{SR}}$, containing $\hat{H}_{\boldsymbol{k}} - \varepsilon_{n\boldsymbol{k}}$, and Chern magnetization $\mathcal{M}_{\text{C}}$, including $2(\varepsilon_{n\boldsymbol{k}} - \mu)$. Within a topologically non-trivial gap, $\mathcal{M}_{SR}$ remains constant, while $\mathcal{M}_C$ varies linearly with $\mu$, with a universal slope $\frac{\partial \mathcal{M}_C}{\partial \mu} = C\frac{e}{h}$ (Fig. 2d).

Under the external magnetic field $\boldsymbol{B}$, the magnetization couples to the field, lowering the total energy. If we assume that the interacting ground state does not undergo qualitative changes apart from a constant energy shift, the change in total energy can be estimated as $\Delta E_{\text{tot}} = -\,\mathcal{M} \cdot \boldsymbol{B}$. In this total energy change, the $\mathcal{M}_{\text{SR}}$ contribution couples to $B$ and directly shifts the bulk band spectrum [18], whereas the $\mathcal{M}_C$ contribution does not affect the bulk spectrum but instead adjusts the edge-mode energies. Among the LAF, QAH, and LPI phases, only the QAH phase possesses a large $\mathcal{M}_{\text{SR}}$, and is therefore more stabilized than the competing LAF and LPI phases under $\boldsymbol{B}$ field. Near the phase boundaries, the total energies of the competing states vary almost linearly with $D$ (Fig. 4d). Since the magnetic correction to $E_{\text{tot}}$ is also linear in $\boldsymbol{B}$, tuning $\boldsymbol{B}$ leads to a linear expansion of the phase boundaries along the $D$ axis (Fig. 4f inset). This behavior is consistent with the experimental trend observed in Fig. 4f.

**Chern insulator state at $D > 0$**

In the main text, we discussed the phase map and associated topological transitions for $D < 0$. Extended Data Fig. 8a shows the longitudinal resistance $R_{xx}(n, D)$ measured for $D > 0$ at $B_a = 0$ T. Unlike the negative $D$ regime, the LAF state apparently evolves directly into LPI state through an intermediate metallic phase. Upon applying $B_a$, a CI phase begins to compete with the metallic state (Extended Data Figs. 8b,c).

To determine $C$ of the field induced CI, we fit the Landau fan diagram of $R_{xx}$ using the Streda formula (Extended Data Fig. 8b). The most dominant Landau level emerges with $C = \pm 5$, and the corresponding $R_{yx}$ reaches the full quantized value $\pm\frac{h}{5e^2}$ for $B_a \geq 0.5$ T. However, the CI gap starts opening already at $B_a = 0.19$ T, as shown in Fig. 4f and Extended Data Fig. 4b and marked by the green stars in Extended Data Figs. 8b,c. A high-resolution map of $B_z^{ac}(n, D)$ at $B_a = 0.25$ T (Extended Data Fig. 8e) shows the tri-striped magnetization pattern as in the QAH state consistent with the appearance of sharp FO transition peaks in $R_{yx}^{ac}$ in Extended Data Fig. 8f.

To further probe the CI state, we acquired $B_z^{ac}(x, y)$ maps as a function of $n$ at fixed $D = 0.155$ V/nm, in the middle of the $D$ range of the gapped state (Extended Data Figs. 8g-l). The evolution of gapped domains closely resembles those of the QAH state at $D < 0$ in Figs. 2g-l: negative differential magnetization domains nucleate first at the right edge of the device, expand across the sample, and eventually shift toward the left edge before disappearing with further doping. From the combination of these observations, we thus conclude that the $C = \pm 5$ CI at $D > 0$ is topologically the same as the QAH state at $D < 0$, which requires, however, a finite $B_a$ to become the ground state. One possible scenario for this asymmetry is that, at $D > 0$, the dielectric screening from $WS_2$ weakens the electron-electron interaction among the valence-band electrons, thereby reducing the stability of the QAH phase.

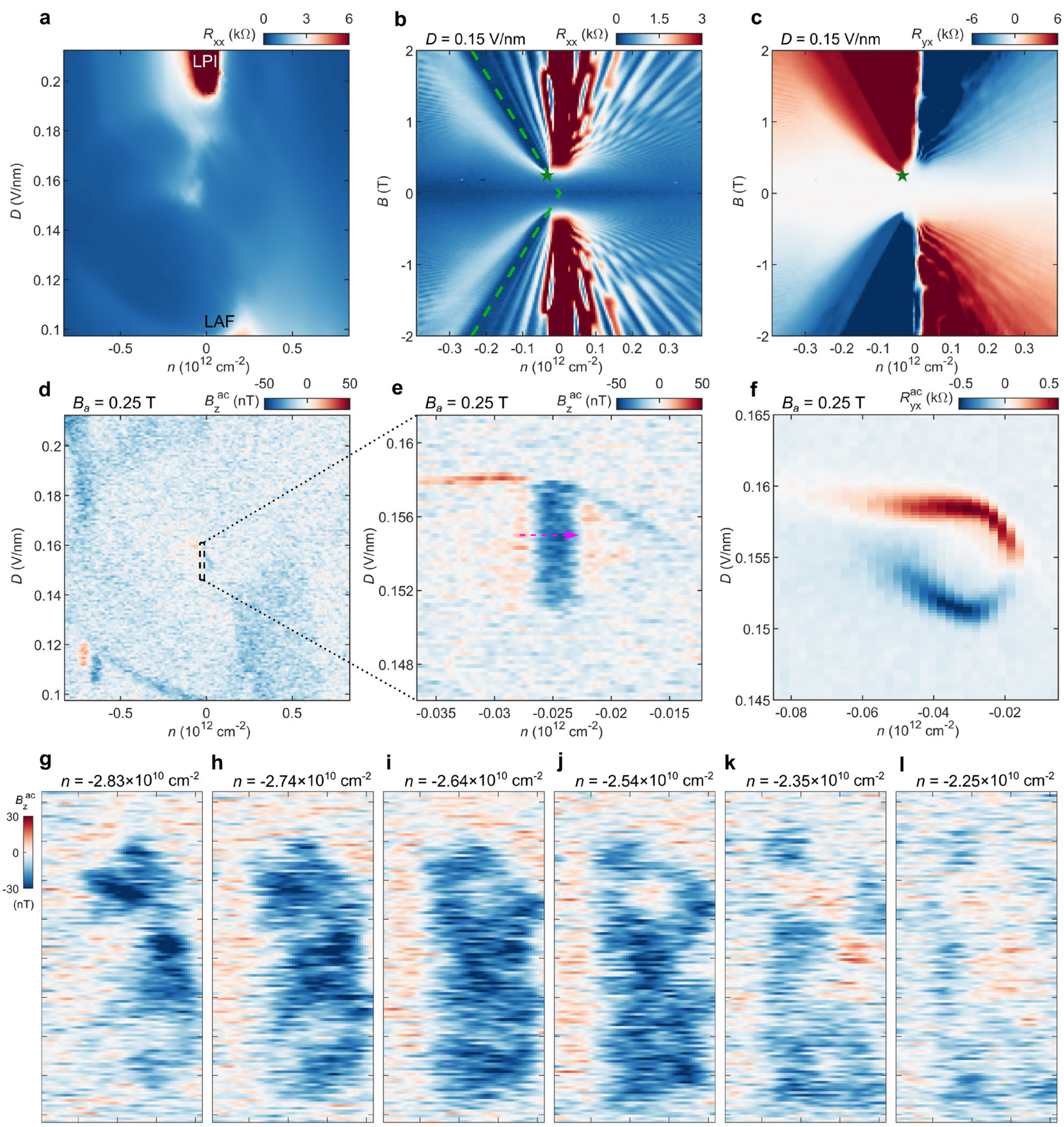

**Extended Data Fig. 8. Chern insulator state at $D > 0$. a,** $R_{xx}(n, D)$ at $B_a = 0$ in the metallic-like region at $D > 0$. **b,c,** Landau fan map of $R_{xx}(n, B_a)$ and $R_{yx}(n, B_a)$ measured at fixed $D = 0.15$ V/nm. Green dashed lines in (b) correspond to the Streda formula fit with $C = \pm 5$. Green stars indicate the onset field of the gapped CI state. **d,** $B_z^{ac}(n, D)$ measured over the same range as in (a) at $B_a = 0.25$ T using $V_{bg}^{ac} = 100$ mV rms. **e,** High resolution map of $B_z^{ac}(n, D)$ measured at $B_a = 0.25$ T within the dashed box in (d) using $V_{bg}^{ac} = 5$ mV rms. **f,** $D^{ac}$-modulated differential Hall resistance $R_{yx}^{ac}(n, D)$ acquired at $B_a = 0.25$ T using purely $D^{ac}$ modulation. **g-l,** $B_z^{ac}(x, y)$ maps a function of $n$ along the magenta dashed arrow in (e) using $V_{bg}^{ac} = 6$ mV rms at $D = 0.155$ V/nm and $B_a = 0.25$ T.

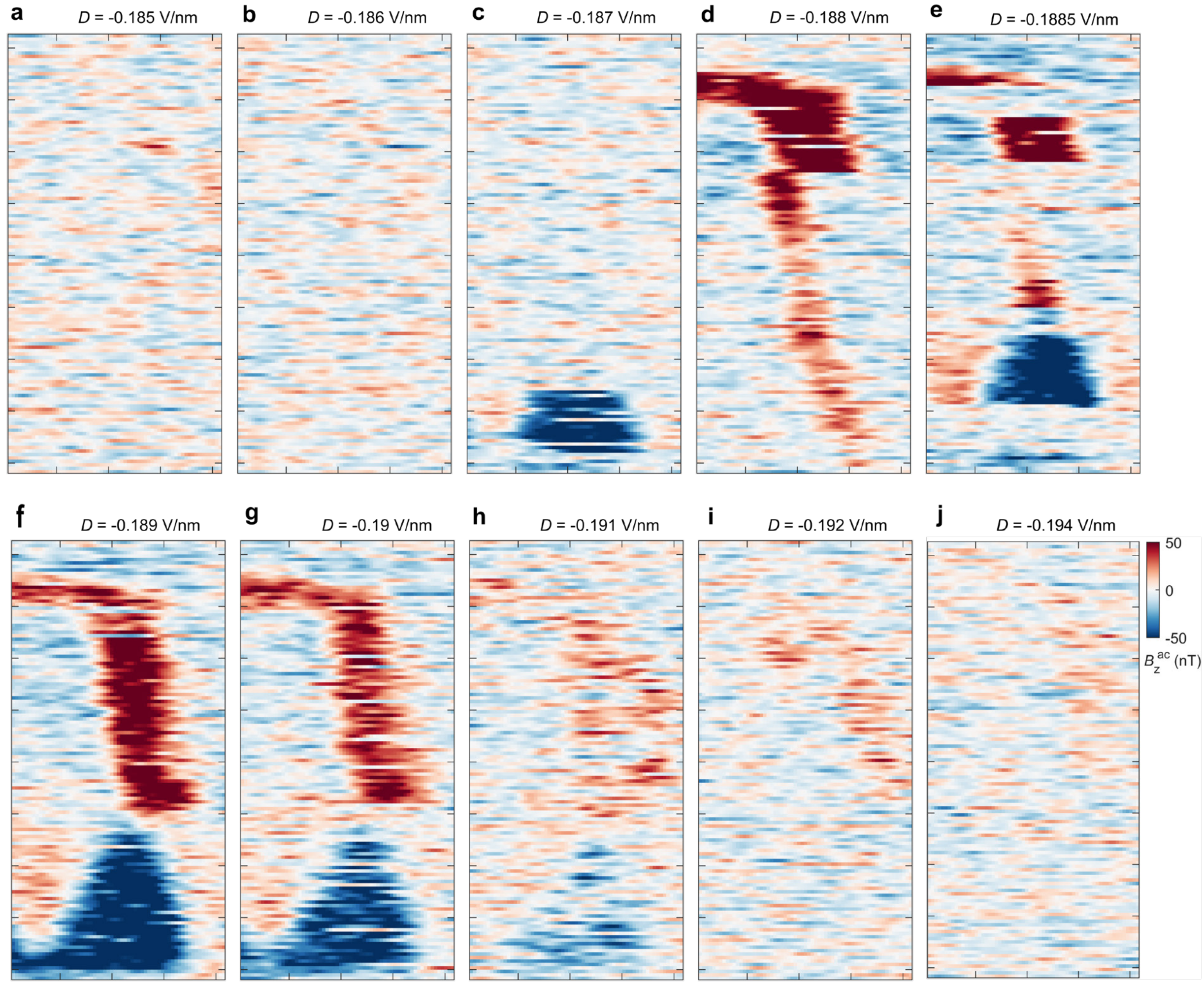

**Extended Data Fig. 9. Displacement field dependence of magnetization fluctuations in M2 phase. a-j,** $B_z^{ac}(x, y)$ images as function of $D$ at $n = -1.2\times10^{10}$ cm$^{-2}$ and applied magnetic field $B_a$ = 12 mT, acquired with $V_{bg}^{ac} = 6$ mV rms along the dashed green line in Fig. 5c. The images reveal fluctuations between opposite polarity magnetic domains.

**Reconstruction of 2D magnetization**

To reconstruct the magnetization from the measured stray-field maps, we employ a self-supervised, untrained physics-informed neural network (UPINN) [18,63,64] that combines a convolutional encoder–decoder architecture with a physics-constrained forward model. The network directly reconstructs the local magnetization from the measured $B_z^{\rm ac}$ without requiring pre-training or ground-truth magnetization data. The reconstructed magnetization is subsequently propagated through the Biot–Savart kernel [65] to generate a simulated stray-field map, $B_z^{ac,sim}$, which is compared with the experimental measurement.

The network adopts a convolutional encoder–decoder architecture consisting of three down-sampling and three up-sampling stages. The encoder employs 64, 128, and 256 feature channels followed by a 512-channel bottleneck, while the decoder mirrors this structure with 256, 128, and 64 feature channels before producing a single-channel magnetization map. Rectified linear unit (ReLU) activations and batch-normalization layers are used throughout the network. The network parameters are optimized by minimizing the difference between the simulated and measured stray-field maps,

$$\mathcal{L} = \left\| B_z^{ac} - B_z^{ac,sim} \right\|^{1/2}$$

Optimization is performed using stochastic gradient descent with momentum for 400 epochs, with an initial learning rate of $10^{-6}$, momentum of 0.4, and a learning-rate decay factor of 0.9 every 10 epochs.

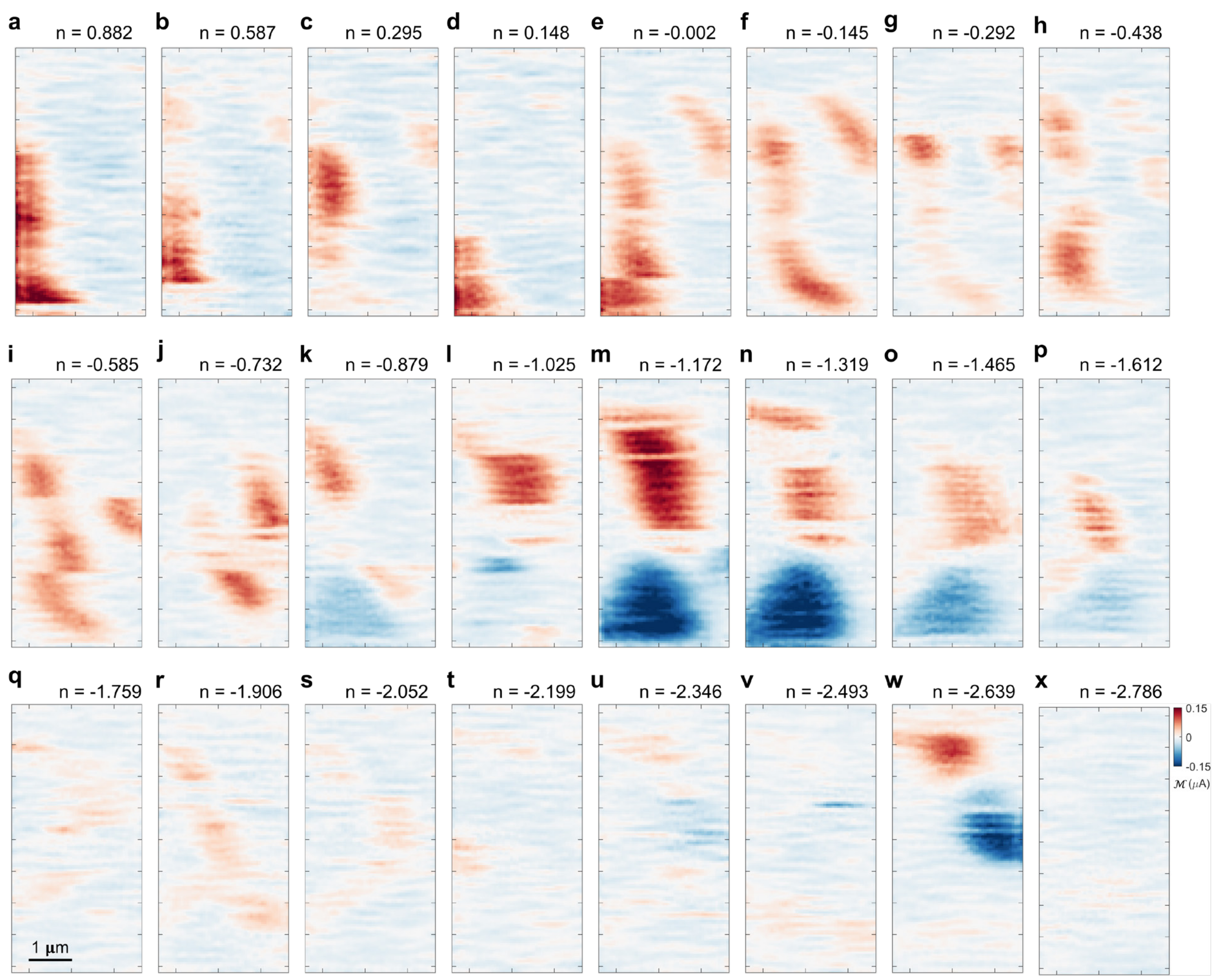

**Extended Data Fig. 10. Differential magnetization in fluctuating M2 phase. a-x,** Reconstructed magnetization maps $\mathcal{M}(x, y)$ from the measured $B_z^{ac}$ signal as a function of $n$ (the indicated values in the titles are units of $10^{10}$ cm$^{-2}$) at fixed $D = -0.188$ V/nm using $V_{bg}^{ac} = 6$ mV rms.

**Supplementary Video 1. Band structure evolution with $U$ in scHF theory.** Left panel: Band structure of the four isospin flavors $\{K\uparrow, K'\uparrow, K\downarrow, K'\downarrow\}$, computed with scHF model at charge neutrality upon varying the displacement potential $U$. The color intensity encodes polarization to top/bottom graphene layers. In the LAF state the $K\downarrow$ and $K'\downarrow$ bands are almost degenerate with a large negative gap that grows with $U$, whereas the positive gap of $K\uparrow$ and $K'\uparrow$ bands shrink with $U$. At FO transition to QAH state, the $K\uparrow$ gap changes layer polarization sign discontinuously, shifting all the bands. In the QAH state all four bands have different gap sizes and relative energies, such that the global gap is determined by $K'\uparrow$ valence and $K\uparrow$ conduction bands. At FO transition to LPI state, the $K'\uparrow$ gap flips sign, shifting all the bands again. In the LPI state all four bands have negative gaps. The four valence bands are almost degenerate, with $K\uparrow$ and $K'\uparrow$ bands having the same gap, which is larger than the gap of $K\downarrow$ and $K'\uparrow$ bands. Right panel: Total energy $E_{tot}$ vs. $U$ of the competing LAF (blue), QAH (red), and LPI (green) states.

**Supplementary Video 2. Band structure evolution with $D$ in continuous topological transition model.** Band structure of the four isospin flavors $\{K\uparrow, K'\uparrow, K\downarrow, K'\downarrow\}$ with increasing the displacement field $D$, computed using a phenomenological tight binding model (Methods). In contrast to scHF theory, the four bands evolve continuously with $K\uparrow$ and $K'\uparrow$ positive band gaps shrinking and $K\downarrow$ and $K'\downarrow$ negative gaps growing with $D$. As $|D|$ increases, the gap of the $K\uparrow$ band (orange) shrinks transitioning from a fully gapped

LAF state to M1 semimetal with overlapping valence and conduction bands with no global gap. The $K\uparrow$ gap then reopens with a positive sign, driving the system into the QAH state. The same process then repeats with $K'\uparrow$ band (cyan) transitioning from QAH state to M2 semimetal. The $K'\uparrow$ gap then reopens with opposite sign, and the system transitions into LPI state with all the gaps having the same sign. In this non-interacting model all the properties, including magnetization and transport, evolve continuously, inconsistent with the experimental findings.

**Supplementary Video 3. Band structure evolution of the QAH state upon doping in scHF theory.** Band structure of the four isospin flavors $\{K\uparrow, K'\uparrow, K\downarrow, K'\downarrow\}$ in the QAH state at $U = -7.6$ meV, computed with scHF model upon hole doping from charge neutrality. The color intensity denotes top/bottom graphene layer polarization. Even though at charge neutrality the $K\downarrow$ and $K'\downarrow$ valence bands are almost degenerate, the hole doping occurs only in the $K'\uparrow$, band, while other three bands remain charge neutral. This yields a symmetry-broken QAH-quarter-metal phase (QAH-1/4M). This selective doping indicates that the sufficiently doped QAH-1/4M phase is not merely a consequence of broken isospin symmetry at charge neutrality, but rather an interacting, Stoner-type phase stabilized under doping. This new quarter-metallic phase is distinct from the previously known LPI-1/4M. This doped QAH-1/4M transitions into the doped LAF-1/4M and three possible LPI-1/4M states through FO transitions (Extended Data Fig. 7).